\documentclass[11pt, onecolumn, a4paper]{article}
\usepackage{amsthm}
\usepackage{amssymb}
\usepackage{amsmath,amstext,amsgen,amsopn,amsbsy,amsxtra,amscd}

\newtheorem{theorem}{Theorem}[section]

\newtheorem{lemma}{Lemma}[section]

\newtheorem{definition}{Definition}[section]

\begin{document}

\title{Analysis of Stochastic Service Guarantees in Communication Networks: A Basic Calculus}

\author{Yuming Jiang\\
Centre for Quantifiable Quality of Service in Communication Systems\\
Department of Telematics\\
Norwegian University of Science and Technology, Norway\\
ymjiang@ieee.org
}
%\date{02 November 2005}

\maketitle

\begin{abstract}
A basic calculus is presented for stochastic service guarantee analysis in communication networks. Central to the calculus are two definitions, {\em maximum-(virtual)-backlog-centric (m.b.c) stochastic arrival curve} and {\em stochastic service curve}, which respectively generalize {\em arrival curve} and {\em service curve} in the deterministic network calculus framework. With m.b.c stochastic arrival curve and stochastic service curve, various basic results are derived under the (min, +) algebra for the general case analysis, which are crucial to the development of stochastic network calculus. These results include (i) superposition of flows, (ii) concatenation of servers, (iii) output characterization, (iv) per-flow service under aggregation, and (v) stochastic backlog and delay guarantees. In addition, to perform independent case analysis, {\em stochastic strict server} is defined, which uses an {\em ideal service process} and an {\em impairment process} to characterize a server. The concept of stochastic strict server not only allows us to improve the basic results (i) -- (v) under the independent case, but also provides a convenient way to find the stochastic service curve of a serve. Moreover, an approach is introduced to find the m.b.c stochastic arrival curve of a flow and the stochastic service curve of a server.\\

\end{abstract}

%\begin{keywords}
%Stochastic network calculus, Stochastic arrival curve, Stochastic service curve, Stochastic strict server, Stochastic quality of service guarantee, Independent case analysis
%\end{keywords}

\section{Introduction}\label{sec_1}
The increasing demand on transmitting multimedia and other real time applications over the Internet has motivated the study of quality of service guarantees. Although these applications are both delay and loss sensitive, they usually can tolerate some delay and loss. As a result, stochastic quality of service guarantees may be well suitable for such applications. In addition, many types of networks only provide stochastic service guarantees.  Examples of such networks are wireless networks and multiaccess networks. In wireless networks, the capacity of a wireless channel may vary with time in a random manner due to channel impairment, contention and other causes. In multiaccess networks such as CSMA (carrier sense multiple access) networks, the server capacity seen by a user also varies over time, which depends largely on the traffic characteristics of other users. Because of this, research on stochastic service guarantees has become critical. In particular, the development of an {\em information theory} for stochastic service guarantee analysis has been identified as a {\em grand challenge} for future networking research \cite{nsf-report03}. Towards it, {\em stochastic network calculus}, the probabilistic version of the {\em (deterministic) network calculus} \cite{Cruz91a}\cite{Cruz91b}\cite{Cruz95}\cite{Chang94}\cite{Boudec98} \cite{Agrawal99} \cite{NetCal}\cite{Chang02}, has been recognized by researchers as a crucial step (e.g. \cite{nsf-report03}\cite{Liebeherr04}).

In recent years, many attempts have been made for the development of a stochastic network calculus, which include \cite{Kurose92}\cite{YS93}\cite{Chang94}\cite{Chang96}\cite{Lee95}\cite{Cruz96}\cite{QK99}\cite{SS00}\cite{gSBB}\cite{Liebeherr-tr02}\cite{Li-tr03}\cite{netcal-tr04}\cite{AF04}\cite{ITC05}\cite{IWQoS05}\cite{CBL05}. However, the following basic properties required by a network calculus have made the stochastic network calculus challenging:
\begin{itemize}
\item[(P.1)] {\bf (Superposition Property)} The superposition of flows can be represented using the same traffic model;
\item[(P.2)] {\bf(Concatenation Property)} The concatenation of servers can be represented using the same server model;
\item[(P.3)] {\bf(Output Characterization)} The output of a flow from a server can be represented using the same traffic model;
\item[(P.4)] {\bf(Per-Flow Service)} The service received by a flow in an aggregate can be represented using the same server model;
\item[(P.5)] {\bf(Service Guarantees)} Stochastic backlog and delay guarantees can be derived.
\end{itemize}
The need of the above summarized properties has been extensively (although separately) discussed in the literature, e.g. (P.1) (P.3) (P.5) in \cite{Kurose92}\cite{YS93}\cite{Chang94}, (P.4) in \cite{QK99}\cite{conf}, and (P.2) in \cite{IWQoS05}\cite{CBL05}. Among these required properties (P.1) -- (P.5), (P.1) is only related to traffic model and (P.2) only to server model, while the other three are related to both. Built upon  (deterministic) {\em arrival curve} for traffic model and (deterministic) {\em service curve} for server mode, the deterministic network calculus developed under the (min, +) algebra \cite{minplus} has all these properties. However, for the stochastic network calculus, to the best of our knowledge, no literature attempt has successfully addressed (P.1) -- (P.5) in all. 

In addition to properties (P.1) -- (P.5), another challenge for developing the stochastic network calculus is to provide approaches to characterizing traffic or service with parameters or models that are known or can be easily obtained. In the deterministic network calculus, token bucket has been used for implementation and description of an arrival curve; Guaranteed Rate server \cite{GV97} and equivalently Latency Rate server \cite{SV98} \cite{Jiang03}, to which many well-known schedulers have been proved to belong (e.g. \cite{Jiang03}\cite{NetCal}), can be used to find the service curve of a server. Among existing attempts for the stochastic network calculus, only a few have focused on this challenge, which include \cite{Chang94}\cite{Chang96}\cite{ITC05} for traffic model and \cite{Li-tr03}\cite{IWQoS05} for server model.  

The third challenge is specific to the stochastic network calculus, which is {\em independent case analysis}. In the deterministic network calculus, dependence between flows and servers is not taken into account since the analysis is based on the worst case. However, for the stochastic network calculus, since traffic processes and service processes are generally stochastic processes, it is natural to analyze cases where these processes are independent so that improved results may be obtained. To date, in the context of stochastic network calculus, although independent case analysis results are available, most of them are for property (P.1) (e.g. \cite{Chang94}\cite{Chang96}) and no independent case analysis result is available for (P.2) -- (P.5) when the server is stochastic. 

Until now, the development of a stochastic network calculus successfully addressing these there  challenges remains an open problem. The main contribution of this paper is to define models and derive results for these challenges. These models and results in all form a basic calculus for stochastic service guarantee analysis. In particular, first, we define a traffic model and a server model. These two models are {\em maximum-(virtual)-backlog-centric (m.b.c) stochastic arrival curve} and {\em stochastic service curve}. The former is a generalization of (deterministic) arrival curve based on its maximum virtual backlog property: the maximum backlog of a virtual single server queue fed with a flow with an arrival curve is upper-bounded. The latter is a generalization of (deterministic) service curve, which is stronger than a previously used stochastic server model \cite{Cruz96}\cite{conf}\cite{netcal-tr04}, called {\em weak stochastic service curve} in this paper. With m.b.c stochastic arrival curve and stochastic service curve, we prove properties (P.1) -- (P.5) under the (min, +) algebra \cite{minplus} for the general case where flows and servers could be dependent. 

Second, we derive properties (P.1) -- (P.5) for the independent case. For this, we define {\em stochastic strict server}, which is based on a simple observation and uses two stochastic processes to characterize a server. These two processes are an {\em ideal service process} and an {\em impairment process}. We show that if the impairment process has a m.b.c stochastic arrival curve, a stochastic strict server has a stochastic service curve. Importantly, the notions of stochastic strict server and impairment process allow us to perform study on (P.2) -- (P.5) for the independent case and obtain results that can be significantly better than those from the (min, +) analysis. 

Third, we prove results that can be used to find the m.b.c stochastic arrival curve of a flow and the stochastic service curve of a server. For this, we focus on the m.b.c stochastic arrival curve of a stochastic process, with which the stochastic service curve of a stochastic strict server can also be derived. We prove that under some general conditions, if a process is $(\sigma(\theta), \rho(\theta))$-upper constrained \cite{Chang94}\cite{Chang96}\cite{Chang00}, it has a m.b.c stochastic arrival curve. Since many widely used processes can be characterized using $(\sigma(\theta), \rho(\theta))$ \cite{Chang94}\cite{Chang96}\cite{Chang00}, they can be readily represented using m.b.c stochastic arrival curve. 

The rest of the paper is organized as follows. In Section \ref{sec_2}, we introduce the network model and derive some preliminary results. In Section \ref{sec_3}, we present the basic stochastic network calculus under the (min, +) algebra, which include the m.b.c stochastic arrival curve traffic model, the stochastic service curve server model, and the basic properties (P.1) -- (P.5). In Section \ref{sec_4}, we define stochastic strict server and prove properties (P.1) -- (P.5) for the independent case. In Section \ref{sec_5}, we show that a $(\sigma(\theta), \rho(\theta))$ process has a m.b.c stochastic arrival curve. Finally we conclude the paper in Section \ref{sec_6}.

\section{Network Model and Preliminaries} \label{sec_2}

In this section, we present the network model, introduce notation, and review and derive some preliminary results that will be used in later analysis.

\subsection{Network Model and Notation}
We consider a lossless communication system which is modeled by various processes. A process is defined to be a function of time $t$, $(t=0, 1, 2, \dots)$. It could count the amount of traffic arriving to some network element, the amount of traffic departing from the network element, the amount of service provided by the network element, or the amount of service failed to be provided by the network element due to some impairment to it.  In this case, we call the process {\em arrival process} denoted by $A(t)$, {\em departure process} denoted by $A^{*}(t)$, {\em service process} denoted by $S(t)$, or {\em impairment process} denoted by $I(t)$ respectively. In this paper, we assume all processes are defined on $t \ge 0 $ and by convention, have zero value at $t=0$, i.e. $A(0)=A^*(0)=S(0)=I(0)=0$. For any $0 \le s \le t$, we denote $A(s,t) \equiv A(t) - A(s)$, $A^{*}(s, t) \equiv A^{*}(t)-A^{*}(s)$, $S(s,t) \equiv S(t) - S(s)$, and $I(s,t)\equiv I(t)-I(s)$.

Wherever necessary, we use subscripts to distinguish different flows, and use superscripts to distinguish different network elements. Specifically, $A_i^{n}$ and $A_i^{n*}$ represent the arrival and departure processes of flow $i$ from network element $n$ respectively, $S_i^{n}$ the service process provided to flow $i$ by the network element, and $I^{n}$ the impairment process suffered by the network element.

For any processes $X(t)$ and $Y(t)$, the following inequalities hold. They can be easily verified. Particularly, for (\ref{i_maxl}), $\sup_{0 \le s \le t} [X(s) + Y(s)] \le \sup_{0 \le s \le t} [X(s) + \sup_{0 \le s \le t}Y(s)] = \sup_{0 \le s \le t} X(s) + \sup_{0 \le s \le t} Y(s)$. (\ref{i_minmg}) can be proved similarly. 
\begin{eqnarray}
\sup_{0 \le s \le t} [X(s) + Y(s)] &\le& \sup_{0 \le s \le t} X(s) + \sup_{0 \le s \le t} Y(s) \label{i_maxl}\\
\inf_{0 \le s \le t} [X(s) - Y(s)] &\ge& \inf_{0 \le s \le t} X(s) - \sup_{0 \le s \le t} Y(s). \label{i_minmg} %\nonumber
\end{eqnarray}

For any two random variables $X$ and $Y$, we say $X$ is stochastically equal to (smaller than; larger than) $Y$, written $X =_{st} (\le_{st}; \ge_{st}) Y$, if $P\{X > x\} = (\le; \ge) P\{Y > x\}$ or $P\{X \le x\} = (\ge; \le) P\{Y \le x\}$ for all $x$ \cite{Stoyan83}\cite{Ross96}.

We denote by ${\mathcal{F}}$ the set of non-negative wide-sense increasing functions, or 
$$
{\mathcal{F}} = \{f(\cdot) : \forall 0 \le x \le y, f(x) \ge 0, f(x) \le f(y)\},
$$
and by ${\mathcal{\bar{F}}}$ the set of non-negative wide-sense decreasing functions, or, 
$$
{\mathcal{\bar{F}}} = \{f(\cdot) : \forall 0 \le x \le y, f(x) \ge 0, f(y) \le f(x)\}.
$$

By definition, $A$, $A^{*}$, $S$ and $I$ belong to ${\mathcal{F}}$. In addition, for any random variable $X$, its distribution function, denoted by $F_X(x) \equiv P\{X \le x\}$, belongs to ${\mathcal{{F}}}$ and its compliment distribution function, denoted by $\bar{F}_X \equiv P\{X>x\}$, belongs to ${\mathcal{\bar{F}}}$.  

We adopt the following operations defined under the (min, +) algebra \cite{minplus}: 
\begin{itemize}
\item The {\em (min, +) convolution} of functions $f$ and $g$ is 
$$
(f \otimes g) (x) = \inf_{0 \le y \le x}[ f(y)+ g(x-y)].
$$
\item The {\em (min, +) deconvolution} of functions $f$ and $g$ is 
$$
(f \oslash g) (x) = \sup_{y \ge 0}[ f(x+y) - g(y)].
$$
\item The {\em pointwise minimum} of functions $f$ and $g$ is
$$
(f \land g) (x) = min[f(x), g(x)].
$$
\item The {\em pointwise maximum} of functions $f$ and $g$ is
$$
(f \lor g) (x) = max[f(x), g(x)].
$$
\end{itemize}
In addition, we shall need the normal convolution for independent case analysis: 
\begin{itemize}
\item The {\em normal convolution} of functions $f$ and $g$ is 
$$
(f \ast g) (x) = \int_{0}^{x} f(x-y) d g(y).
$$
\end{itemize}

For function $f$ in ${\mathcal{F}}$, we set $f(x) = f(0)$ for any $x<0$. For function $f$ in ${\mathcal{\bar{F}}}$, we also set $f(x) = f(0)$ for any $x<0$. For the (min, +) convolution of functions in ${\mathcal{\bar{F}}}$, we set $f\otimes g (x) = +\infty$ for any $x <0$. For the normal convolution of functions in ${\mathcal{\bar{F}}}$, we set $(f \ast g) (x) = 0$ for any $x <0$.

For ease of exposition, we adopt 
$$[x]_{1} \equiv min[x, 1] \quad \textrm{and} \quad (x)^+ \equiv max[x, 0].$$

For later analysis, we need the following result that is a special case of literature results for one function of two random variables (e.g. see p. 141 of \cite{Papoulis91}), which can also be easily verified. 
\begin{lemma}\label{l_plus}
Consider random variable $X$. For any $x \ge 0$, $P\{(X)^+ >x\} = P\{X > x\}$. 
\end{lemma}

\subsection{Background on Service Guarantee Analysis} \label{sec_22}

For service guarantee analysis of a system, we are interested in two quantities: {\em backlog} and {\em delay} defined as \cite{Cruz95}\cite{Chang00}\cite{NetCal}:
\begin{itemize} 
\item The backlog $B(t)$ in the system at time $t$ is \\ $B(t) = A(t) - A^{*}(t)$; 
\item The delay $D(t)$ at time $t$ is \\ $D(t) = \inf\{d \ge 0 : A(t) \le A^{*}(t+d)\}$.
\end{itemize} 

In the context of network calculus for deterministic service guarantee analysis, the (deterministic) {\em arrival curve} traffic model and the (deterministic) {\em service curve} server model are the most fundamental and important concepts. Their definitions are as follows: 

\begin{definition} \label{d_ac}
A flow is said to have a (deterministic) arrival curve $\alpha \in {\mathcal{F}}$ iff for all $0 \le s \le t$, there holds (e.g. \cite{NetCal})
\begin{equation} \label{e_ac}
A(s, t) \le \alpha (t-s).
\end{equation}
\end{definition}

\begin{definition} \label{d_sc}
A server is said to provide a (deterministic) service curve $\beta \in {\mathcal{F}}$ to its arrival $A$, iff for all $t \ge 0$, its departure $A^{*}$ satisfies (e.g. \cite{NetCal})
\begin{equation} \label{e_sc}
A^{*}(t) \ge A\otimes \beta(t). 
\end{equation}
\end{definition}

Inspired by the ideas behind arrival curve and service curve, several probabilistic versions of them have been proposed in the literature for the stochastic network calculus \cite{YS93}\cite{Lee95}\cite{Cruz96}\cite{SS00}\cite{gSBB}\cite{Liebeherr-tr02}\cite{Li-tr03}\cite{netcal-tr04}\cite{AF04}\cite{ITC05}\cite{CBL05}. 

For traffic models, they can be generalized into two models, which are called in this paper {\em t.a.c stochastic arrival curve} and {\em v.b.c stochastic arrival curve}. In particular, a flow is said to have a {\em traffic-amount-centric (t.a.c) stochastic arrival curve} $\alpha \in {\mathcal{F}}$ with bounding function $f \in {\mathcal{\bar{F}}}$, denoted by $A \sim_{ta} \langle f, \alpha \rangle $, iff, for all $0 \le s \le t$ and all $x \ge 0$, there holds
\begin{equation} \label{e_sac1}
P\{ A(s, t) - \alpha (t-s) > x \} \le f(x).
\end{equation}
A flow is said to have a {\em virtual-backlog-centric (v.b.c) stochastic arrival curve} $\alpha \in {\mathcal{F}}$ with bounding function $f \in {\mathcal{\bar{F}}}$, denoted by $A \sim_{vb} \langle f, \alpha \rangle $, iff for all $t \ge 0$ and all $x \ge 0$, there holds
\begin{equation} \label{e_sac2}
P\{ \sup_{0\le s \le t}[A(s, t) - \alpha (t-s)] > x \} \le f(x).
\end{equation}

For server models, most of them belong to what we shall call {\em weak stochastic service curve}. Particularly, a server $S$ is said to prove a {\em weak stochastic service curve} $\beta \in {\mathcal{F}}$ with bounding function $g \in {\mathcal{\bar{F}}}$, denoted by $S \sim_{ws} \langle g, \beta \rangle$, iff for all $t \ge 0$ and all $x \ge 0$, there holds
\begin{equation} \label{e_wssc}
P\{A \otimes \beta(t) - A^{*}(t) > x \} \le g(x).
\end{equation}

It can be easily verified that the EBB (exponentially bounded burstiness) model in \cite{YS93}, the SBB (stochastically bounded burstiness) model in \cite{SS00}, and the effective envelope or statistical envelope in \cite{Liebeherr-tr02}\cite{Li-tr03}\cite{CBL05} are special cases of t.a.c stochastic arrival curve (\ref{e_sac1}). The stochastic smoothness constraint traffic models in \cite{Cruz96}, the gSBB (generalized stochastically bounded burstiness) in \cite{gSBB} \cite{ITC05}, and the traffic model used in \cite{netcal-tr04}\cite{AF04} belong to v.b.c stochastic arrival curve (\ref{e_sac2}). For server models, the EBF (exponentially bounded fluctuation) model in \cite{Lee95}, the stochastic service constraint server models in \cite{Cruz96} and the effective service curve or statistical service curve in \cite{Liebeherr-tr02}\cite{Li-tr03}\cite{CBL05} can be easily mapped to weak stochastic service curve (\ref{e_wssc}). In addition, the server model defined in \cite{netcal-tr04} and used in \cite{conf}\cite{ITC05} is a special case of weak stochastic service curve.

\subsection{Properties of ${\mathcal{\bar{F}}}$, $\land$, $\lor$, and $\otimes$} 

It has been proved that $({\mathcal{{F}}}, \land, \otimes)$ is a complete dioid \cite{minplus}, (which is defined to have all the properties listed in Lemma \ref{l_dio} below), with zero function $\epsilon$ and identity function $e$ where $\epsilon(x)= +\infty$ for all $x \ge 0$, and $e(x)=0$ if $x = 0$ and otherwise $+\infty$ \cite{minplus}\cite{Chang02}\cite{NetCal}. 

For $({\mathcal{\bar{F}}}, \land, \otimes)$, the following result shows it also is a complete dioid with zero function $\bar{\epsilon}$ and identity function $\bar{e}$ where $\bar{\epsilon}(x) = +\infty$ for all $x \ge 0$ and $\bar{e}(x) = 0$ for all $x \ge 0$.

\begin{lemma}\label{l_dio}
$({\mathcal{\bar{F}}}, \land, \otimes)$ is a complete dioid having the following properties:
\begin{itemize}
\item[(i)] {Closure property:} $\forall f, g \in {\mathcal{\bar{F}}}$, $f \land g \in {\mathcal{\bar{F}}}$; $f \otimes g \in {\mathcal{\bar{F}}}$.

\item[(ii)] {Associativity:} $\forall f, g \in {\mathcal{\bar{F}}}$, $(f \land g) \land h = f \land (g \land h)$; $(f \otimes g) \otimes h = f \otimes (g \otimes h)$.

\item[(iii)] {Commutativity:} $\forall f, g \in {\mathcal{\bar{F}}}$,  $f \land g = g \land f$; $f \otimes g = g \otimes f$.

\item[(iv)] {Distributivity: } $\forall f, g, h \in {\mathcal{\bar{F}}}$,  $(f \land g) \otimes h = (f \otimes h) \land (g \otimes h)$. 

\item[(v)] {Zero element:} $\forall f \in {\mathcal{\bar{F}}}$, $f \land \bar{\epsilon} = f$.

\item[(vi)] {Absorbing zero element:} $\forall f \in {\mathcal{\bar{F}}}$, $f \otimes \bar{\epsilon} = \bar{\epsilon} \otimes f = \bar{\epsilon}$.

\item[(vii)] {Identity element:} $\forall f \in {\mathcal{\bar{F}}}$, $f \otimes \bar{e} = \bar{e} \otimes f = f$.

\item[(viii)] {Idempotency of addition:} $\forall f \in {\mathcal{\bar{F}}}$, $f \land f = f$.
\end{itemize}
\end{lemma}

\begin{proof} For (i), if $0 \le x \le y$, $f(x) \ge f(y) \ge 0$ and $g(x) \ge g(y) \ge 0$. Hence, $
(f \land g) (x) = min [f(x), g(x)] \ge min [f(y), g(y)] = (f \land g) (y).$ In addition, $(f\otimes g)(x)= \inf_{0 \le z \le x}[f(z) + g(x-z)] \ge \inf_{0 \le z \le x}[f(z) + g(y-z)] \ge \inf_{0 \le z \le y}[f(z) + g(y-z)] = (f\otimes g)(y)$. Furthermore, for any $x \ge 0$, it can be verified that $(f \land g)(x) \ge 0$ and $(f \otimes g)(x) \ge 0$.

For (ii)--(iv), their proofs are identical to those for $({\mathcal{{F}}}, \land, \otimes)$. 

For (v), (vi) and (viii), they can be easily verified to hold. For (vii), since $f$ is wide-sense decreasing, $f \otimes \bar{e}(x) = \inf_{0 \le y \le x}f(y) = f(x)$ and $\bar{e} \otimes f (x) = \inf_{0 \le y \le x} f(x-y) = f(x)$. 
\end{proof}

In addition, we have the following properties for functions in ${\mathcal{\bar{F}}}$:

\begin{lemma}\label{l_mon}
$\forall f_1, f_2, g_1, g_2 \in {\mathcal{\bar{F}}}$, 
\begin{itemize}
\item[(ix)] {Comparison: } $f_1 \land f_2 \le f_1 \lor f_2 \le f_1 \otimes f_2$;
\item[(x)] {Monotonicity: } If $f_1 \le g_1$ and $f_2 \le g_2$, then $f_1 \otimes f_2 \le g_1 \otimes g_2$; $f_1 \land f_2 \le g_1 \land g_2$; $f_1 \lor f_2 \le g_1 \lor g_2$.
\end{itemize}
\end{lemma}

\begin{proof}
We shall only prove (ix) and the others can be verified easily. 

Let us first prove $f_1 \le f_1 \otimes f_2$. By definition, $(f_1 \otimes f_2) (x) = \inf_{0 \le y \le x} [f_1 (y) + f_2(x-y)]$. Since $f_2(x-y) \ge 0$, we get $(f_1 \otimes f_2) (x) \ge \inf_{0 \le y \le x} [f_1 (y)]$. In addition, since $f_1$ is wide sense decreasing, $f_1 (y) \ge f_1(x)$ for all $(0 \le) y \le x$. Hence, $\inf_{0 \le y \le x} [f_1 (y)] = f_1(x)$ and $(f_1 \otimes f_2) (x) \ge f_1 (x)$. 

Similarly, we can prove $(f_1 \otimes f_2) (x) \ge f_2 (x)$. Hence, $(f_1 \otimes f_2) (x) \ge (f_1\lor f_2) (x)$. In addition, it is trivially true that $(f_1 \lor f_2)(x) \ge (f_1 \land f_2)(x)$ and then (ix) follows.
\end{proof}

For stochastic service guarantee analysis, we shall rely on compliment distribution functions of random variables, which are functions in ${\mathcal{\bar{F}}}$. For such functions, we have the following result: 

\begin{lemma} \label{l_rsum}
For any random variables $X$ and $Y$, and $\forall x \ge 0$, there holds, 
\begin{equation}\label{e_rsum0}
\bar{F}_{ X+Y}(x) \le (\bar{F}_{X} \otimes \bar{F}_{Y})(x),
\end{equation}
and, if $\bar{F}_{X}(x) \le f(x)$ and $\bar{F}_{Y}(x) \le g(x)$, where $f, g \in {\mathcal{\bar{F}}}$, then 
\begin{equation} \label{e_rsum}
P\{ X+Y > x \} \le  (f \otimes g)(x).
\end{equation}
\end{lemma}
\begin{proof}
For any $y \ge 0$, $\{ X+Y > x\} \cap \{X \le y\} \cap \{Y \le x -y\} = \phi$, where $\phi$ denotes the null set. We then have $\{ X+Y > x\} \subset \{X > y\} \cup \{Y > x -y\}$ and hence $P \{ X+Y > x\} \le P\{X > y\} + P\{Y > x -y\}$. Since this derivation holds for all $y$, $(0 \le y \le x)$, we get $P \{ X+Y > x\} \le \inf_{0 \le y \le x }[P\{X > y\} + P\{Y > x -y\}]$, which is (\ref{e_rsum0}). Then, with the monotonicity property of $\otimes$, (\ref{e_rsum}) follows from (\ref{e_rsum0}).
\end{proof}

\section{A Basic Stochastic Network Calculus} \label{sec_3}
In this section, we introduce a new generalization of arrival curve, a new generalization of service curve, and their basic properties (P.1) -- (P.5) for stochastic service guarantee analysis.

\subsection{Traffic Model}

The idea of the new generalization of the arrival curve traffic model is based on the {\em triplicity principle} of arrival curve stated by Lemma \ref{l_acd} below:

\begin{lemma} \label{l_acd}
The following statements are equivalent:
\begin{itemize}
\item[(i)] $\forall 0 \le s \le t$, $A(s,t) \le \alpha(t-s) + x$ for all $x \ge 0$;
\item[(ii)] $\forall t \ge 0$, $\sup_{0 \le s \le t}[A(s,t)- \alpha(t-s)] \le x$ for all $x \ge 0$;
\item[(iii)] $\forall t \ge 0$, $\sup_{0 \le s \le t}\sup_{0 \le u \le s}[A(u,s)- \alpha(s-u)] \le x$ for all $x \ge 0$,
\end{itemize}
where $\alpha \in {\mathcal{F}}$.
\end{lemma}
\begin{proof}
It is trivially true that $A(s,t) - \alpha(t-s) \le \sup_{0 \le s \le t}[A(s,t)- \alpha(t-s)]$,  from which, (ii) implies (i). In addition
\begin{eqnarray}
&& \sup_{0 \le s \le t}[A(s,t)- \alpha(t-s)] \nonumber\\
&\le& \sup_{0 \le s \le t}\sup_{s \le v \le t}[A(s,v)- \alpha(v-s)] \nonumber\\
&=& \sup_{0 \le v \le t}\sup_{0 \le s \le v}[A(s,v)- \alpha(v-s)] \nonumber\\
&=& \sup_{0 \le s \le t}\sup_{0 \le u \le s}[A(u,s)- \alpha(s-u)] \label{e_acd}
\end{eqnarray}
with which, (iii) implies (ii).

For (i) $\rightarrow$ (ii), it holds since $A(s,t) - \alpha(t-s) \le x$ for all $0 \le s \le t$. For (ii) $\rightarrow$ (iii), $\sup_{0 \le s \le t}\sup_{0 \le u \le s}[A(u,s)- \alpha(s-u)] \le \sup_{0 \le s \le t}[x] = x$. 

Hence (i), (ii) and (iii) are equivalent.
\end{proof}

By the definition of arrival curve, the right hand side of $A(s,t) \le \alpha(t-s) + x$ in Lemma \ref{l_acd}.(i) defines an arrival curve $\alpha(t-s) + x$ or the traffic amount $A(s,t)$ is upper-bounded by $\alpha(t-s) + x$. In addition, let us construct a virtual single server queue system that is initially empty, fed with the same traffic $A$, and has service curve $\alpha$ making $A^*(t) \ge A \otimes \alpha(t)$. Then, the backlog in the virtual system is upper-bounded by $A(t) -A^*(t) \le \sup_{0 \le s \le t}[A(s,t)- \alpha(t-s)] \le x$, and the maximum backlog up-to-date in the virtual system is upper-bounded by $\sup_{0 \le s \le t} [A(s) - A^*(s)] \le \sup_{0 \le s \le t}\sup_{0 \le u \le s}[A(u,s)- \alpha(s-u)] \le x$. Calling Lemma \ref{l_acd}.(i) the {\em traffic amount property} of arrival curve, Lemma \ref{l_acd}.(ii) its {\em virtual backlog property}, and Lemma \ref{l_acd}.(iii) its {\em maximum virtual backlog property}, Lemma \ref{l_acd} states that the three properties of arrival curve are equivalent. It is in this sense we call Lemma \ref{l_acd} the {\em triplicity principle} of arrival curve.

Based on the traffic amount property and virtual backlog property of arrival curve, two probabilistic versions of arrival curve have been proposed, which, as discussed earlier in Section \ref{sec_22}, are respectively {\em traffic-amount-centric (t.a.c) stochastic arrival curve} and {\em virtual-backlog-centric (v.b.c) stochastic arrival curve}.

We now introduce a new probabilistic version of arrival curve, which is based on its maximum virtual backlog property. 

\begin{definition} \label{d_mbc}
A flow is said to have a {\em maximum-(virtual)-backlog-centric (m.b.c) stochastic arrival curve} $\alpha \in {\mathcal{F}}$ with bounding function $f \in {\mathcal{\bar{F}}}$, denoted by $A \sim_{mb} \langle f, \alpha \rangle $, iff for all $t \ge 0$ and all $x \ge 0$, there holds
\begin{equation} \label{e_sac3}
P\{ \sup_{0\le s \le t}\sup_{0\le u \le s}[A(u, s) - \alpha (s-u)] > x \} \le f(x).
\end{equation}
\end{definition}

The following lemma states that (deterministic) arrival curve is a special case of m.b.c stochastic arrival curve.

\begin{lemma}\label{l_dsa}
A flow has a (deterministic) arrival curve $\alpha$, if and only if it has a m.b.c stochastic arrival curve $A \sim_{mb} \langle 0, \alpha \rangle$.
\end{lemma}
\begin{proof}
For the ``only if'' part, since the flow has an arrival curve $\alpha$, we have $A(u, s) \le \alpha(s-u)$, or $A(u, s) - \alpha(s-u) \le 0$ for all $0 \le u \le s$. Hence $P\{ \sup_{0\le s \le t}\sup_{0\le u \le s}[A(u, s) - \alpha (s-u)] > x \} = 0$ for all $x \ge 0$.

For the ``if'' part, it is given that $P\{ \sup_{0\le s \le t}\sup_{0\le u \le s}[A(u, s) - \alpha (s-u)] > 0 \} = 0$. In other words, there holds  $\sup_{0\le s \le t}\sup_{0\le u \le s}[A(u, s) - \alpha (s-u)] \le 0$. Note that if there would exist some $0 \le u_0 \le s_0 \le t$ making $A(u_0, s_0) - \alpha (s_0-u_0) >0$, we would have $\sup_{0\le s \le t}\sup_{0\le u \le s}[A(u, s) - \alpha (s-u)] > 0$. Hence, we must have $A(u,s) \le \alpha(s-u)$ for all $0 \le u \le s \le t$. This ends the proof. 
\end{proof}

Based on the definitions of t.a.c stochastic arrival curve, v.b.c stochastic arrival curve and m.b.c stochastic arrival curve, since $A(s,t) - \alpha(t-s) \le \sup_{0 \le s \le t}[A(s,t)- \alpha(t-s)] \le \sup_{0 \le s \le t}\sup_{0 \le u \le s}[A(u,s)- \alpha(s-u)]$, the following relationship between them is immediately obtained.

\begin{lemma}\label{l_mva}
$A \sim_{mb} \langle f, \alpha \rangle $ $\longrightarrow$ $A \sim_{vb} \langle f, \alpha \rangle $ $\longrightarrow$  $A \sim_{ta} \langle f, \alpha \rangle $, where $X \longrightarrow Y$ means $X$ implies $Y$.
\end{lemma}

It is worth highlighting that while for arrival curve, both Lemma \ref{l_acd}.(i) $\rightarrow$ .(ii) $\rightarrow$ .(iii) and Lemma \ref{l_acd}.(i) $\leftarrow$ .(ii) $\leftarrow$ .(iii) hold, for stochastic arrival curve, we generally do not have $A \sim_{mb} \langle f, \alpha \rangle $ $\leftarrow$ $A \sim_{vb} \langle f, \alpha \rangle $ $\leftarrow$  $A \sim_{ta} \langle f, \alpha \rangle $.

\subsection{Server Model}

For defining the new generalization of the service curve server model, we explore the following {\em duality principle} of service curve. 

\begin{lemma}\label{l_scd}
For any $x \ge 0$, $A \otimes \beta(t) - A^{*}(t) \le x$ for all $t \ge 0$, if and only if $\sup_{0 \le s \le t}[A \otimes \beta(s) - A^{*}(s)] \le x$ for all $t \ge 0$, where $\beta \in {\mathcal{F}}$.
\end{lemma}
\begin{proof}
For the ``if'' part, it holds trivially since $A \otimes \beta(t) - A^{*}(t) \le \sup_{0 \le s \le t}[A \otimes \beta(s) - A^{*}(s)]$. For the ``only if'' part, since $A \otimes \beta(t) - A^{*}(t) \le x$ for all $t \ge 0$, $\sup_{0 \le s \le t}[A \otimes \beta(s) - A^{*}(s)] \le \sup_{0 \le s \le t}[x] = x$.
\end{proof}

By the definition of service curve, it is clear that the first part of Lemma \ref{l_scd} defines a service curve $\beta(t) - x$. Lemma \ref{l_scd} states that if a server provides service curve $\beta(t) - x$, then there holds $\sup_{0 \le s \le t}[A \otimes \beta(s) - A^{*}(s)] \le x$, and vice versa. In this sense, we call Lemma \ref{l_scd} the {\em duality principle} of service curve.

Inspired by the first part of Lemma \ref{l_scd}, a probabilistic version of service curve has been proposed and studied in the literature, which is called {\em weak stochastic service curve} as discussed earlier in Section \ref{sec_22}.

In the following, we introduce a new probabilistic extension of service curve based on the second part of Lemma \ref{l_scd}. 

\begin{definition} \label{d_ssc}
A server $S$ is said to prove a {\em stochastic service curve} $\beta \in {\mathcal{F}}$ with bounding function $g \in {\mathcal{\bar{F}}}$, denoted by $S \sim_{sc} \langle g, \beta \rangle $, iff for all $t \ge 0$ and all $x \ge 0$, there holds
\begin{equation} \label{e_ssc}
P\{ \sup_{0 \le s \le t}[ A \otimes \beta(s) - A^{*}(s)] > x \} \le g(x).
\end{equation}
\end{definition}

The following lemma states that (deterministic) service curve is a special case of stochastic service curve, which can be proved using the same approach as for Lemma \ref{l_dsa}.

\begin{lemma} \label{l_dss}
A server has a (deterministic) service curve $\beta$, if and only if it has a stochastic service curve $S \sim_{sc} \langle 0, \beta \rangle$.
\end{lemma}

Comparing the definitions of {\em weak stochastic service curve} and {\em stochastic service curve}, since $A \otimes \beta(t) - A^{*}(t) \le \sup_{0 \le s \le t}[ A \otimes \beta(s) - A^{*}(s)]$, the following relationship between them is obtained.

\begin{lemma} \label{l_ws}
$S \sim_{sc} \langle g, \beta \rangle $ $\longrightarrow$ $S \sim_{ws} \langle g, \beta \rangle$, where $X \longrightarrow Y$ means $X$ implies $Y$.
\end{lemma}

It is also worth highlighting that while for service curve, the two parts in Lemma \ref{l_scd} are equivalent, for stochastic service curve, we only have $S \sim_{sc} \langle g, \beta \rangle $ $\longrightarrow$ $S \sim_{ws} \langle g, \beta \rangle$. This is also the reason why we call the probabilistic extension of service curve based on the first part of Lemma \ref{l_scd} {\em weak} stochastic service curve.

\subsection{Basic Properties (P.1) -- (P.5)}

Having defined {\em m.b.c stochastic arrival curve} and {\em stochastic service curve}, we now prove properties (P.1) -- (P.5) under the (min, +) algebra for the general case where flows and servers could be dependent.

\begin{theorem} \label{t_sup}
{\bf (P.1: Superposition)} Consider $N$ flows with arrival processes $A_i(t)$, $i=1, \dots, N$, respectively. Let $A(t)$ denote the aggregate arrival process, or $A(t) = \sum_{i=1}^{N}A_i(t)$. If $\forall i, A_i \sim_{mb} \langle f_i, \alpha_i \rangle$, then $A \sim_{mb} \langle f, \alpha \rangle$ where $f(x)=f_1 \otimes \cdots \otimes f_{N}(x)$ and $\alpha (t)= \sum_{i=1}^{N}\alpha_i(t)$.
\end{theorem}
\begin{proof}
We only prove the case of $N=2$, from which the proof can be easily extended to $N > 2$ through iteration.

Following the same approach as in the proof of (\ref{i_maxl}), we get $\sup_{0 \le s \le t}\sup_{0 \le u \le s}[A_1(u,s) - \alpha_1(s-u) + A_2(u,s) - \alpha_2(s-u)] \le \sup_{0 \le s \le t}\sup_{0 \le u \le s}[A_1(u,s) - \alpha_1(s-u)] + \sup_{0 \le s \le t}\sup_{0 \le u \le s}[A_2(u,s) - \alpha_2(s-u)]$. Then, 
\begin{eqnarray}
&& \sup_{0 \le s \le t}\sup_{0 \le u \le s}[A(u,s) - \alpha(s-u)] \nonumber\\
&=& \sup_{0 \le s \le t}\sup_{0 \le u \le s}[A_1(u,s) - \alpha_1(s-u) \nonumber \\
&& + A_2(u,s) - \alpha_2(s-u)] \nonumber \\
&\le& \sup_{0 \le s \le t}\sup_{0 \le u \le s}[A_1(u,s) - \alpha_1(s-u)] \nonumber\\
&& + \sup_{0 \le s \le t}\sup_{0 \le u \le s}[A_2(u,s) - \alpha_2(s-u)] \nonumber\\
&\le& \left(\sup_{0 \le s \le t}\sup_{0 \le u \le s}[A_1(u,s) - \alpha_1(s-u)]\right)^+ \nonumber\\
&& + \left(\sup_{0 \le s \le t}\sup_{0 \le u \le s}[A_2(u,s) - \alpha_2(s-u)]\right)^+ \label{ts_e}
\end{eqnarray} 
with which, Lemma \ref{l_rsum}, and the definition of m.b.c stochastic arrival curve, the theorem is proved.
\end{proof}

%\begin{remark}
{\bf Remark:} For t.a.c stochastic arrival curve and v.b.c stochastic arrival curve, their superposition properties have also been proved (e.g. \cite{SS00}\cite{gSBB} \cite{netcal-tr04}\cite{ITC05}).
%\end{remark}

\begin{theorem}\label{t_con}
{\bf (P.2: Concatenation)} Consider a flow passing through a network of $N$ nodes in tandem. If each node $n(=1, 2, \dots, N)$ provides stochastic service curve $S^n \sim_{sc} \langle g^n, \beta^n \rangle$ to its input, then the network guarantees to the flow a stochastic service curve $S \sim_{sc} \langle g, \beta \rangle$ with 
\begin{eqnarray}
\beta(t) &=& \beta^1 \otimes \beta^2 \otimes \cdots \otimes \beta^N (t) \nonumber \\
g (x) &=& g^1 \otimes g^2 \otimes \cdots \otimes g^N (x). \nonumber 
\end{eqnarray}
\end{theorem}

\begin{proof}
We shall only prove the two-node case, from which the proof can be easily extended to the $N$-node case. For the two-node case, the departure of the first node is the arrival to the second node, so, $A^{1*}(t) = A^2(t)$. In addition, the arrival to the network is the arrival to the first node, or $A(t)=A^{1}(t)$, and the departure from the network is the departure from the second node, or $A^{*}(t)=A^{2*}(t)$, where $A(t)$ and $A^{*}(t)$ denote the arrival process to and departure process from the network respectively. We then have,
\begin{eqnarray}
&& \sup_{0 \le s \le t}[A\otimes \beta^1 \otimes \beta^2 (s) - A^*(s)] \nonumber\\
&=& \sup_{0 \le s \le t}[(A^1\otimes \beta^1) \otimes \beta^2(s) - A^{2*}(s)] \label{tc_1}.
\end{eqnarray}

Now let us consider any $s$, $(0 \le s \le t)$, for which we get,
\begin{eqnarray}
&& [(A^1\otimes \beta^1) \otimes \beta^2(s) - A^{2*}(s)] \nonumber \\
&& - \sup_{0 \le u \le t}[A^1 \otimes \beta^1 (u) - A^{1*}(u)] \nonumber \\
&& - \sup_{0 \le u \le t}[A^2 \otimes \beta^2 (u)-A^{2*}(u)] \nonumber \\
&\le& (A^1\otimes \beta^1) \otimes \beta^2(s) - A^{2*}(s) \nonumber \\ 
&& - \sup_{0 \le u \le s}[A^1 \otimes \beta^1 (u) - A^{1*}(u)] \nonumber \\
&& - \sup_{0 \le u \le s}[A^2 \otimes \beta^2 (u)-A^{2*}(u)] \nonumber \\
&=& (A^1\otimes \beta^1) \otimes \beta^2(s) \nonumber \\
&& - \sup_{0 \le u \le s}[A^1 \otimes \beta^1 (u) - A^{2}(u)] \nonumber \\
&& - \sup_{0 \le u \le s}[A^2 \otimes \beta^2 (u)+ A^{2*}(s) -A^{2*}(u)] \nonumber \\
&\le& \inf_{0 \le u \le s}[A^1\otimes \beta^1 (u) + \beta^2(s-u)] \nonumber \\
&& - \sup_{0 \le u \le s}[A^1 \otimes \beta^1 (u) - A^{2}(u)] \nonumber \\
&& - \sup_{0 \le u \le s}[A^2 \otimes \beta^2 (u)] \nonumber\\
&\le& \inf_{0 \le u \le s}[(A^1\otimes \beta^1 (u) + \beta^2(s-u)) \nonumber \\
&& - (A^1 \otimes \beta^1 (u) - A^{2}(u))] - \sup_{0 \le u \le s}[A^2 \otimes \beta^2 (u)] \nonumber\\
&=& \inf_{0 \le u \le s} [A^{2}(u) + \beta^2(s-u)] - \sup_{0 \le u \le s}[A^2 \otimes \beta^2 (u)] \nonumber \\
&=& A^2 \otimes \beta^2 (s) - \sup_{0 \le u \le s}[A^2 \otimes \beta^2 (u)] \nonumber \\
&\le& 0 \label{tc_2}.
\end{eqnarray}

Applying (\ref{tc_2}) to (\ref{tc_1}), we obtain
\begin{eqnarray}
&& \sup_{0 \le s \le t}[A\otimes \beta^1 \otimes \beta^2 (s) - A^*(s)] \nonumber \\
&\le& \sup_{0 \le u \le t}[A^1 \otimes \beta^1 (u) - A^{1*}(u)] \nonumber \\
&& + \sup_{0 \le u \le t}[A^2 \otimes \beta^2 (u)-A^{2*}(u)] \label{tc_e} 
%\\&\le& \left(\sup_{0 \le u \le t}[A^1 \otimes \beta^1 (u) - A^{1*}(u)]\right)^+ \nonumber \\
%&& + \left(\sup_{0 \le u \le t}[A^2 \otimes \beta^2 (u)-A^{2*}(u)]\right)^+ \nonumber
\end{eqnarray}
with which, since both nodes provide stochastic service curve to their input, the theorem follows from Lemma \ref{l_rsum} and the definition of stochastic service curve.
\end{proof}

%\begin{remark}
{\bf Remark:} In deriving (\ref{tc_2}), we have proved $[(A^1\otimes \beta^1) \otimes \beta^2(s) - A^{2*}(s)] \le \sup_{0 \le u \le s}[A^1 \otimes \beta^1 (u) - A^{1*}(u)] + \sup_{0 \le u \le s}[A^2 \otimes \beta^2 (u)-A^{2*}(u)]$ for all $s \ge 0$. However, if we want to prove the concatenation property for weak stochastic service curve, we need to prove $[(A^1\otimes \beta^1) \otimes \beta^2(s) - A^{2*}(s)] \le [A^1 \otimes \beta^1 (s) - A^{1*}(s)] + [A^2 \otimes \beta^2 (s)-A^{2*}(s)]$ for all $s \ge 0$, which is difficult to obtain and does not hold in general. This explains why weak stochastic service curve does not have property (P.2). 
%\end{remark}

\begin{theorem}\label{t_out}
{\bf (P.3: Output Characterization)} Consider a server fed with a flow. If the server provides stochastic service curve $S \sim_{sc} \langle g, \beta \rangle$ to the flow and the flow has m.b.c stochastic arrival curve $A \sim_{mb} \langle f, \alpha \rangle$, then the departure process of the flow from the server has a m.b.c stochastic arrival curve $A^{*} \sim_{mb} \langle f^{*}, \alpha^{*} \rangle$ with 
\begin{eqnarray}
\alpha^{*}(t) &=& \alpha \oslash \beta (t) \nonumber \\
f^{*}(x) &=&  f \otimes g (x). \nonumber
\end{eqnarray}
\end{theorem}
\begin{proof}
The departure up to time $t$ cannot exceed the arrival in $[0, t]$, or $A^{*}(t) \le A(t)$. We now have,
\begin{eqnarray}
&& \sup_{0 \le s \le t}\sup_{0 \le u \le s}[A^{*}(u,s) - \alpha^{*}(s-u)] \nonumber \\
&=& \sup_{0 \le s \le t}\sup_{0 \le u \le s}[A^{*}(s) -A^{*}(u) - \alpha^{*}(s-u)] \nonumber \\
&\le& \sup_{0 \le s \le t}\sup_{0 \le u \le s}[A(s) -A^{*}(u) - \alpha^{*}(s-u)] \nonumber \\
&=&  \sup_{0 \le s \le t}\sup_{0 \le u \le s}[A(s) - A\otimes \beta(u) - \alpha^{*}(s-u) \nonumber \\
&& + A\otimes \beta(u) - A^{*}(u) ] \nonumber \\
&\le& \sup_{0 \le s \le t}\sup_{0 \le u \le s}[A(s) - A\otimes \beta(u) - \alpha^{*}(s-u)] \nonumber \\
&& + \sup_{0 \le u \le t}[A\otimes \beta(u) - A^{*}(u) ] \label{to_1}
\end{eqnarray}
in which, step (\ref{to_1}) follows from a similar approach as in proving (\ref{i_maxl}). 

In addition, 
\begin{eqnarray}
&&\sup_{0 \le u \le s}[A(s) - A\otimes \beta(u) - \alpha^{*}(s-u)] \nonumber \\
&=&\sup_{0 \le u \le s}[A(s) - \inf_{0 \le v \le u}[A(v) + \beta(u-v)] - \alpha \oslash \beta(s-u)] \nonumber\\
&=&\sup_{0 \le u \le s}\sup_{0 \le v \le u}[A(s) - A(v) - \beta(u-v) - \alpha\oslash \beta(s-u)] \nonumber\\
%&\le&\sup_{0 \le u \le s}\sup_{0 \le v \le u}[A(s,v) - \beta(u-v) - [\alpha(s-v) - \beta(u-v)]] \label{to_2}\\
&\le& \sup_{0 \le u \le s}\sup_{0 \le v \le u}[A(v, s) - \alpha(s-v)] \label{to_2} \\
%&\le& \sup_{0 \le u \le s}\sup_{0 \le v \le u}[A(v, s)- \alpha(s-v)] + \sup_{s \ge 0}[\alpha(s) - \beta(s)] \label{to_2} \\
&=& \sup_{0 \le v \le s}\sup_{v \le u \le s}[A(v, s)- \alpha(s-v)] \nonumber \\
&=& \sup_{0 \le v \le s}[A(v, s)- \alpha(s-v)] \label{to_3}
\end{eqnarray}
where step (\ref{to_2}) holds because by the definition of $\oslash$, $\alpha\oslash \beta(s-u) = \sup_{w \ge 0}[\alpha(s-u + w) - \beta(w)] \ge \alpha(s-v) - \beta(u-v)$ by taking $w = u-v$. Applying (\ref{to_3}) to (\ref{to_1}), we get
\begin{eqnarray}
&& \sup_{0 \le s \le t}\sup_{0 \le u \le s}[A^{*}(u,s) - \alpha^{*}(s-u)] \nonumber \\
&\le& \sup_{0 \le s \le t}\sup_{0 \le u \le s}[A(u, s)- \alpha(s-u)] \nonumber \\
&& + \sup_{0 \le s \le t}[A\otimes \beta(s) - A^{*}(s) ] \label{to_e}
%\\&\le& \left(\sup_{0 \le s \le t}\sup_{0 \le u \le s}[A(u, s)- \alpha(s-u)]\right)^+ \nonumber \\
%&& + \left(\sup_{0 \le s \le t}[A\otimes \beta(s) - A^{*}(s) ]\right)^+ \nonumber
\end{eqnarray}
with which, since $S \sim_{sc} \langle g, \beta \rangle$ and $A \sim_{mb} \langle f, \alpha \rangle$, or $P\{\sup_{0 \le s \le t}[A\otimes \beta(s) - A^{*}(s) ] >x\} \le g(x)$ and $P\{\sup_{0 \le s \le t}\sup_{0 \le u \le s}[A\otimes \beta(u) - A^{*}(u) ] >x\} \le f(x)$, the theorem follows from Lemma \ref{l_rsum}.
\end{proof}

%\begin{remark}
{\bf Remark:} Following similar steps, it can be proved that property (P.3) also holds if the arrival is modeled with v.b.c stochastic arrival curve and the server is modeled with stochastic service curve (e.g. \cite{IWQoS05}). However, it is difficult to prove property (P.3) for other combinations of stochastic arrival curves and stochastic service curves.
%\end{remark}

\begin{theorem}\label{t_per}
{\bf (P.4: Per-Flow Service)} Consider a server fed with a flow $A$ that is the aggregation of two constituent flows $A_1$ and $A_2$. Suppose the server provides stochastic service curve $S \sim_{sc} \langle g, \beta \rangle$ to the aggregate flow $A$. If flow $A_2$ has m.b.c stochastic arrival curve $A_2 \sim_{mb} \langle f_2, \alpha_2 \rangle$ and  $\beta'_1 \in {\mathcal{F}}$, then the server guarantees to flow $A_1$ stochastic service curve $S_1 \sim_{sc} \langle g'_1, \beta'_1 \rangle$, where,
\begin{eqnarray}
%\beta'_1(t) &=& \left( \beta(t) - \alpha_2(t) \right)^+ \nonumber \\
\beta'_1(t) &=& \beta(t) - \alpha_2(t)  \nonumber \\
g'_1(x) &=& g\otimes f_2(x). \nonumber
\end{eqnarray}
\end{theorem}
\begin{proof}
For the departure, there holds $A^*(t) = A^*_1(t) + A^*_2(t)$. In addition, we have $A^*(t) \le A(t)$, $A^*_1(t) \le A_1(t)$, and $A^*_2(t) \le A_2(t)$. We now have for any $s \ge 0$, 
\begin{eqnarray}
&& A_1 \otimes (\beta - \alpha_2)(s) - A^*_1(s) \nonumber\\
%&=& \inf_{0 \le u \le s} [A_2(u) + \beta(s-u)-\alpha(s-u)] - A^*(s) + A^*_1(s) \nonumber\\
&=& \inf_{0 \le u \le s} [A(u) + \beta(s-u)-\alpha_2(s-u) -A_2(u)] \nonumber\\
&& - A^*(s) + A^*_2(s) \nonumber\\
&\le& [A\otimes \beta(s) - A^*(s)] + A_2(s) \nonumber\\&& 
- \inf_{0 \le u \le s}[A_2(u) + \alpha_2(s-u)] \nonumber \\
&=& [A\otimes \beta(s) - A^*(s)] \nonumber\\&&
+ \sup_{0 \le u \le s}[A_2(u,s) - \alpha_2(s-u)] \label{tp_1}.
\end{eqnarray}

Hence,
\begin{eqnarray}
&& \sup_{0 \le s \le t}[A_1 \otimes (\beta - \alpha_2)(s) - A^*_1(s)] \nonumber\\
&\le& \sup_{0 \le s \le t}[A\otimes \beta(s) - A^*(s)] \nonumber\\
&& + \sup_{0 \le s \le t}\sup_{0 \le u \le s}[A_2(u,s) - \alpha_2(s-u)] \label{tp_e}
%\\&\le& \left(\sup_{0 \le s \le t}[A\otimes \beta(s) - A^*(s)]\right)^+ \nonumber\\
%&& + \left(\sup_{0 \le s \le t}\sup_{0 \le u \le s}[A_2(u,s) - \alpha_2(s-u)]\right)^+ \nonumber
\end{eqnarray}
with which, $S \sim_{sc} \langle g, \beta \rangle$, $A_2 \sim_{mb} \langle f_2, \alpha_2 \rangle$ and the definition of stochastic service curve, the theorem follows.
\end{proof}

%\begin{remark}
{\bf Remark:} Based on (\ref{tp_1}), it can be proved that property (P.4) also holds if the traffic model is  v.b.c stochastic arrival curve and the server model is weak stochastic service curve (e.g. \cite{netcal-tr04}). However, for other combinations of the three stochastic arrival curve models and the two stochastic service curve models, this property is difficult to obtain.
%\end{remark}

\begin{theorem}\label{t_ser}
{\bf (P.5: Service Guarantees)} Consider a server fed with a flow $A$. If the server provides stochastic service curve $S \sim_{sc} \langle g, \beta \rangle$ to the flow and the flow has m.b.c stochastic arrival curve $A \sim_{mb} \langle f, \alpha \rangle$, then 
\begin{itemize}
\item[(i)] The backlog $B(t)$ of the flow in the server at time $t$ satisfies: for all $t \ge 0$ and all $x \ge 0$,
\begin{equation}
P \{B(t) > x\} \le f\otimes  g(x+ \inf_{s \ge 0}[\beta(s) - \alpha(s)]);
\end{equation}
\item[(ii)] The delay $D(t)$ of the flow in the server at time $t$ satisfies: for all $t \ge 0$ and all $x \ge 0$,
\begin{equation}\label{d1}
P \{D(t) > x\} \le f\otimes g(\inf_{s \ge 0}[\beta(s) - \alpha(s-x)]).
\end{equation}
\end{itemize}
\end{theorem}

\begin{proof}
For the backlog, by definition $B(t)=A(t) - A^*(t)$. In addition, 
\begin{eqnarray}
&& A(t) - A^*(t) = A(t) - A\otimes \beta(t) + A\otimes \beta(t) - A^*(t) \nonumber \\
%&\le& \sup_{0 \le s \le t}[A(t) - A(t-s) - \beta(s)] + \sup_{0 \le s \le t}[A\otimes \beta(t) - A^*(t)] \nonumber \\
&\le& \sup_{0 \le s \le t}[A(t-s,t) - \alpha(s) +\alpha(s) - \beta(s)] \nonumber\\
&& + \sup_{0 \le s \le t}[A\otimes \beta(t) - A^*(t)] \nonumber\\
&\le& \sup_{0 \le s \le t}[A(t-s,t) - \alpha(s)] \nonumber\\
&& + \sup_{0 \le s \le t}[A\otimes \beta(t) - A^*(t)] + \sup_{s \ge 0}[\alpha(s)-\beta(s)] \nonumber\\
&=& \sup_{0 \le s \le t}[A(s,t) - \alpha(t-s)] \nonumber\\
&& + \sup_{0 \le s \le t}[A\otimes \beta(t) - A^*(t)] - \inf_{s \ge 0}[\beta(s) - \alpha(s)] \nonumber\\
&\le& \sup_{0 \le s \le t}\sup_{0 \le u \le s}[A(u,s) - \alpha(s-u)] \nonumber\\
&& + \sup_{0 \le s \le t}[A\otimes \beta(t) - A^*(t)] - \inf_{s \ge 0}[\beta(s) - \alpha(s)] \label{tb_e}
%\\&\le& \left(\sup_{0 \le s \le t}\sup_{0 \le u \le s}[A(u,s) - \alpha(s-u)]\right)^+ \nonumber\\
%&& + \left(\sup_{0 \le s \le t}[A\otimes \beta(t) - A^*(t)]\right)^+ - \inf_{s \ge 0}[\beta(s) - \alpha(s)] \nonumber
\end{eqnarray}
in which step (\ref{tb_e}) follows from (\ref{e_acd}). With (\ref{tb_e}), if $x+\inf_{s \ge 0}[\beta(s) - \alpha(s)] <0$, the first part holds trivially since we have set $f \otimes g(y) = +\infty$ for any $y <0$. If $x+\inf_{s \ge 0}[\beta(s) - \alpha(s)] \ge 0$, it follows from $A \sim_{mb} \langle f, \alpha \rangle$, $S \sim_{sc} \langle g, \beta \rangle$, Lemma \ref{l_plus} and Lemma \ref{l_rsum}.

For the delay, by definition, $D(t) = \inf\{ d \ge 0 : A(t) \le A^{*}(t+d)\}$, which implies that, for any $x \ge 0$, if $D(t) > x$, there must be $A(t) > A^{*}(t+x)$, since otherwise if there would be $A(t) \le A^{*}(t+x)$ and we would have $D(t) \le x$ which contradicts the condition $D(t) > x$. In other words, if event $\{D(t) > x\}$ happens, event $\{A(t) > A^{*}(t+x)\}$ must happen, or $\{D(t) > x\} \subset \{A(t) > A^{*}(t+x)\}$ and $P\{D(t) > x\} \le P\{A(t) > A^{*}(t+x)\}$ \cite{Cruz96}. Following similar steps in (\ref{tb_e}), we obtain
\begin{eqnarray}
&& A(t) - A^*(t+x) \nonumber \\
&=& A(t) - A\otimes \beta(t+x) + A\otimes \beta(t+x)- A^*(t+x) \nonumber\\
&=& \sup_{0 \le s \le t+x}[A(t) - A(s) - \alpha(t-s) + \alpha(t-s) \nonumber\\
&&  - \beta(t+x-s)] + A\otimes \beta(t+x)- A^*(t+x) \label{td_eao}\\
&\le& \sup_{0 \le s \le t+x}[A(t) - A(s) - \alpha(t-s)] \nonumber\\
&& + \sup_{0 \le s \le t+x}[\alpha(t-s) - \beta(t+x-s)] \nonumber\\
&& + A\otimes \beta(t+x)- A^*(t+x) \label{td_ea00}\\
%&\le& \left(\sup_{0 \le s \le t}[A(t) - A(s) - \beta(t+x-s)]\right)^+ \nonumber\\
%&& + A\otimes \beta(t+x)- A^*(t+x) \nonumber\\
%&\le& \left(\sup_{0 \le s \le t}[A(t) - A(s) - \alpha(t-s) + \alpha(t-s) - \beta(t+x-s)]\right)^+ \nonumber\\
%&& + A\otimes \beta(t+x)- A^*(t+x) \nonumber\\
%&\le& \sup_{0 \le s \le t}[A(t) - A(s)] - \beta(t+x) \nonumber\\
%&&+ A\otimes \beta(t+x)- A^*(t+x) \label{td_ea00}\\
&\le& \sup_{0 \le s \le t+x}[A(t) - A(s) - \alpha(t-s)]  \nonumber\\
&& + \sup_{u \ge 0}[\alpha(u-x) - \beta(u)] \nonumber\\
&& + A\otimes \beta(t+x)- A^*(t+x) \nonumber\\ %\label{td_eao}
&\le& \sup_{0 \le s \le t}\sup_{0 \le u \le s}[A(u,s) - \alpha(s-u)] \nonumber\\
&& + \sup_{0 \le s \le t+x}[A\otimes \beta(s) - A^*(s)] \nonumber \\&& 
- \inf_{s \ge 0}[\beta(s) - \alpha(s-x)] \label{td_ea} 
%\\&\le& \left(\sup_{0 \le s \le t}\sup_{0 \le u \le s}[A(u,s) - \alpha(s-u)]\right)^+ \nonumber\\
%&& + \left(\sup_{0 \le s \le t+x}[A\otimes \beta(s) - A^*(s)]\right)^+ \nonumber \\&& 
%- \inf_{s \ge 0}[\beta(s) - \alpha(s-x)] \nonumber
\end{eqnarray}
where step (\ref{td_ea}) holds because $\sup_{0 \le s \le t+x}[A(t) - A(s) - \alpha(t-s)] = max[\sup_{0 \le s \le t}[A(t) - A(s) - \alpha(t-s)], \sup_{t < s \le t+x}[A(t) - A(s) - \alpha(t-s)]]$, and $\sup_{0 \le s \le t}[A(t) - A(s) - \alpha(t-s)] \le \sup_{0 \le s \le t}\sup_{0 \le u \le s}[A(u,s) - \alpha(s-u)]$ as proved in (\ref{e_acd}), and $\sup_{t < s \le t+x}[A(t) - A(s) - \alpha(t-s)] \le -\alpha(0)$ since we have let $\alpha(x) = \alpha(0)$ for all $x <0$ and $-\alpha(0) \le \sup_{0 \le s \le t}\sup_{0 \le u \le s}[A(u,s) - \alpha(s-u)]$.

With (\ref{td_ea}), we get
\begin{eqnarray}
&& P\{D(t) > x\} \nonumber\\
&\le& P \{ X_a + Y_s \ge \inf_{s \ge 0}[\beta(s) - \alpha(s-x)]\} \label{td_e}
%\\&\le& P \{ (X_a)^+ + (Y_s)^+ \ge \inf_{s \ge 0}[\beta(s) - \alpha(s-x)]\} \nonumber
\end{eqnarray}
where $X_a$ and $X_s$ are random variables with $X_a = \sup_{0 \le s \le t}\sup_{0 \le u \le s}[A(u,s) - \alpha(s-u)]$ and $Y_s = \sup_{0 \le s \le t+x}[A\otimes \beta(s) - A^*(s)]$. Finally, since $A \sim_{mb} \langle f, \alpha \rangle$, $P\{X_a >y\} \le f(y)$ for all $y \ge 0$; since $S \sim_{sc} \langle g, \beta \rangle$, $P\{X_s >y\} \le g(y)$ for all $y \ge 0$; then, the delay part follows similarly from (\ref{td_e}) and Lemma \ref{l_rsum}. 
\end{proof}

%\begin{remark}
{\bf Remark:} (i) Literature results have also shown that property (P.5) holds when the traffic model is v.b.c stochastic arrival curve (e.g. \cite{netcal-tr04}\cite{AF04}). However, if the traffic model is t.a.c stochastic arrival curve, a difficulty will be encountered in proving property (P.5) (e.g. see \cite{Li-tr03}), which is, it is difficult to derive $P\{\sup_{0 \le s \le t}[A(s, t) - \alpha(t-s)]>x \}$ given $P\{A(s, t) - \alpha(t-s)>x\}$. 

(ii) Note that if the server is deterministic providing deterministic service curve $\beta$, some possibly better result for delay can be derived. In this case, $A\otimes \beta(t+x)- A^*(t+x) \le 0$. Since we are studying event $\{A(t) - A^{*}(t+x) >0\}$, for the first term in the right hand side of (\ref{td_eao}), we shall only need to consider $s$ with $0 \le s \le t$, since for $t < s \le t+x$, the whole right hand side of (\ref{td_eao}) is less than or equal to 0. Consequently, (\ref{td_ea00}) becomes $A(t) - A^{*}(t+x) \le \sup_{0 \le s \le t}[A(t) - A(s) - \alpha(t-s)] + \sup_{0 \le s \le t}[\alpha(t-s) - \beta(t+x-s)] = \sup_{0 \le s \le t}[A(t) - A(s) - \alpha(t-s)] - \inf_{s \ge 0}[\beta(s+x) - \alpha(s)]$. Eventually, the delay guarantee becomes
\begin{eqnarray}
P \{D(t) > x\} &\le& f\otimes g(\inf_{s \ge 0}[\beta(s+x) - \alpha(s)]) \label{d2}
\end{eqnarray}
where $g(x)=0$ for the deterministic server. To demonstrate the difference between (\ref{d1}) and (\ref{d2}), let us suppose $\beta(t)= r t$, the arrival is also deterministic with $\alpha(t) = \rho t + \sigma$, and $\rho <r$. From (\ref{d1}), we can conclude that the delay is bounded by $D(t) \le \sigma / \rho$, while from (\ref{d2}), $D(t) \le \sigma / r$ which is a tighter bound than the former. Nevertheless, when the server is stochastic, since we do not have $A\otimes \beta(t+x)- A^*(t+x) \le 0$, it is generally difficult to get (\ref{d2}) for the stochastic delay guarantee.\\
%\end{remark}

\subsection{Comparison and Other Related Work}

Table \ref{tbl} summarizes the properties that are provided by the combination of a traffic model, chosen from t.a.c, v.b.c and m.b.c stochastic arrival curves, and a server model, chosen from weak stochastic service curve and stochastic service curve, without any additional constraints on the traffic model or the server model. In Section \ref{sec_22}, we have discussed that under the context of network calculus, most traffic models used in the literature \cite{YS93}\cite{Cruz96}\cite{SS00}\cite{gSBB}\cite{Liebeherr-tr02}\cite{Li-tr03}\cite{netcal-tr04}\cite{AF04}\cite{ITC05}\cite{IWQoS05}\cite{CBL05} belong to t.a.c and v.b.c stochastic arrival curve, and most server models \cite{Lee95}\cite{Cruz96}\cite{Liebeherr-tr02}\cite{Li-tr03}\cite{netcal-tr04}\cite{AF04}\cite{CBL05} belong to weak stochastic service curve. Table \ref{tbl} shows that without additional constraints, these works can only support part of the five required properties for the stochastic network calculus. In contrast, with m.b.c stochastic arrival curve and stochastic service curve, all these properties have been proved in this section.

\begin{table} [htb] %{table*}
%\linespread{1.0}
\caption{Properties provided by a combination of traffic model and server model}
\begin{center}
\begin{tabular}{|c|c|c|}
\hline  & weak st. service curve & st. service curve \\\hline\hline
t.a.c st. arrival curve & (P.1) & (P.1), (P.3) \\\hline
v.b.c st. arrival curve & (P.1), (P.4), (P.5) & (P.1) -- (P.3), (P.5) \\\hline
m.b.c st. arrival curve & (P.1), (P.4), (P.5) & (P.1) -- (P.5) \\\hline
\end{tabular}
\end{center}
\label{tbl}
\end{table}

Note that with some additional constraints on the bounding functions in these models, one combination may have more properties among (P.1) -- (P.5) than those listed in the table. In \cite{netcal-tr04}\cite{CBL05}, a constraint is imposed or implied on the bounding function of a t.a.c or v.b.c stochastic arrival curve and on the bounding function of a weak stochastic service curve. This constraint, first suggested in \cite{SS00}, is that the bounding function belongs to a specific subset of $\mathcal{\bar{F}}$ that consists of all functions in $\mathcal{\bar{F}}$, whose $n$th-fold integration still belongs to the subset for any $n \ge 1$. Under this constraint, the unlisted properties among (P.1) -- (P.5) can be proved for the combination of t.a.c or v.b.c stochastic arrival curve and weak stochastic service curve. However, the formulation usually is much more complex and the proof needs an intermediate step directly related to stochastic service curve  (e.g. see \cite{CBL05} and its Lemma 1). In addition, the corresponding properties of the deterministic network calculus can be recovered from these properties only {\em almost surely} \cite{CBL05}. In contrast, this paper requires no additional constraint on the bounding function of a m.b.c stochastic arrival curve or a stochastic service curve, and all properties (P.1) -- (P.5) are proved directly from their definitions. In addition, all the corresponding properties of the deterministic network calculus can be {\em exactly} recovered from Theorems \ref{t_sup} -- \ref{t_ser}. Also, given the same constraint on the bounding function, similar results as in \cite{netcal-tr04}\cite{CBL05} can be obtained from the analysis in this section. Moreover, this paper proves properties (P.1) -- (P.5) for independent case analysis in the next section, which are missing in previous works based on the concepts of the various stochastic arrival curves and stochastic service curves. 

Also note that Table \ref{tbl} only provides a comparison on the basic properties supported by a combination of the three types of stochastic arrival curves and the two types of stochastic service curves. While we believe they cover a wide range of traffic models and server models proposed and studied in the literature as discussed in Section \ref{sec_22} and above, there are other types of traffic and server models that are not covered by them. 

One type uses a sequence of random variables to stochastically bound the arrival process \cite{Kurose92} or the service process \cite{QK99}. Similar properties as (P.1), (P.3), (P.4) and (P.5) have been studied \cite{Kurose92}\cite{QK99}. These studies generally need the independence assumption. Under this type of traffic and service models, several problems remain open, which are out of the scope of this paper. One is the concatenation property (P.2), another is the general case analysis and the third is researching/designing approaches to map known traffic and service characterizations to the required sequences of random variables. 

Another type is built upon moments or moment generating functions. This type was initially used for traffic \cite{Chang94}\cite{Knightly97} and has also been extended to service \cite{Chang00}\cite{WN03}. Independence assumption is generally required between arrival and service processes. Extensive study has been conducted for deriving the characteristics of a process under this type of model from some known characterization of the process \cite{Chang94}\cite{Chang96}\cite{Chang00}. Main open problems for this type are the concatenation property (P.2) and the general case analysis. Although these problems are out of the scope of this paper, we prove in Section \ref{sec_5} results that relate the moment generating function model to the proposed m.b.c stochastic arrival curve and stochastic service curve. These results will allow to further relate known traffic / service characterization to the proposed traffic and service models in this paper.

\section{Independent Case Analysis} \label{sec_4} 
In the previous section, various results for stochastic service guarantee analysis have been derived under the (min, +) algebra. These results are obtained without considering the dependence condition between flows and servers. In this section, independent case analysis is performed and properties (P.1) -- (P.5) are proved when flows and servers are independent. 

We start with a lemma that is followed by a simple example demonstrating the importance of the independent case analysis. 

\begin{lemma} \label{l_rsumi}
For nonnegative random variables $X$ and $Y$, if they are independent and $\bar{F}_{X}(x) \le f(x)$ and $\bar{F}_{Y}(x) \le g(x)$, where $f, g \in {\mathcal{\bar{F}}}$, then, for all $x \ge 0$, 
\begin{equation} \label{e_rsumi}
P\{ X+Y > x \} \le 1- (\bar{f} \ast \bar{g})(x)
\end{equation}
where $\bar{f}(x) = 1- [f(x)]_1$ and $\bar{g}(x)=1-[g(x)]_1$.
\end{lemma}
\begin{proof}
For independent random variables $X$ and $Y$, it is well known that $F_{X+Y} = \int_{-\infty}^{+\infty} F_{X}(x-y) d F_{Y}(y)$. Since $X$ and $Y$ are nonnegative, $F_X(x) = 0$ and $F_{Y}(x)=0$ for all $x <0$. Hence, $F_{X+Y} = \int_{0}^{x} F_{X}(x-y) d F_{Y}(y) = F_{X} \ast F_{Y} (x)$. Notice that $F_X$, $F_Y$, $\bar{f}$ and $\bar{g}$ are wide sense increasing, $\bar{f} \le F_X$ and $\bar{g} \le F_Y$, and the convolution operation is commutative. Then
\begin{eqnarray}
F_{X} \ast F_{Y} (x) &=& \int_{0}^{x} F_X(x-y) d F_Y(y) \nonumber\\
&\ge& \int_{0}^{x} \bar{f}(x-y)  d F_Y(y) \nonumber \\
&=& \int_{0}^{x} F_Y(x-y) d \bar{f}(y) \nonumber \\
&\ge& \int_{0}^{x} \bar{g} (x-y) d \bar{f}(y) \nonumber \\
&=& (\bar{f} \ast \bar{g})(x), \nonumber
\end{eqnarray}
with which and $P\{ X+Y > x \} = \bar{F}_{X+Y} = 1 - F_{X} \ast F_{Y}$, the lemma is proved.
\end{proof}

{\em Example:} In Lemma \ref{l_rsum}, it has been proved that $P\{ X+Y > x \} \le (f \otimes g) (x)$. If $X$ and $Y$ are independent, we then have two bounds for $P\{ X+Y > x \}$, which are (\ref{e_rsum}) and (\ref{e_rsumi}). Suppose $f(x)=g(x)= e^{-x}$. With Lemma \ref{l_rsum}, we obtain
$$
P\{ X+Y > x \} \le 2 e^{-x/2},
$$
and with Lemma \ref{l_rsumi}, we get
$$
P\{ X+Y > x \} \le (1+x)e^{-x}.
$$
It can be easily verified that the latter bound obtained from Lemma \ref{l_rsumi} is (much) better than the former bound from Lemma \ref{l_rsum}.

The above example shows that by considering the independent condition between two random variables, significant improvement may be obtained for the result.

\subsection{The Difficulty}
From the above example, we expect that when flows and servers are independent in a system, possibly much better results or tighter bounds can be obtained for properties (P.1) -- (P.5). However, except for the superposition property (P.1), it is not straightforward to prove properties (P.2) -- (P.5) for the independent case. 

The difficulty relates to the definitions of m.b.c stochastic arrival curve and stochastic service curve. Particularly, the first two terms on the right hand side of (\ref{tc_e}), \ref{to_e}), (\ref{tp_e}), (\ref{tb_e}) and (\ref{td_ea}) are generally dependent no matter whether the flow(s) and the server(s) are independent. For example, in the following inequality duplicating (\ref{to_e}),
\begin{eqnarray}
&& \sup_{0 \le s \le t}\sup_{0 \le u \le s}[A^{*}(u,s) - \alpha^{*}(s-u)] \nonumber \\
&\le& \sup_{0 \le s \le t}\sup_{0 \le u \le s}[A(u, s)- \alpha(s-u)] + \nonumber \\
&& \sup_{0 \le s \le t}[A\otimes \beta(s) - A^{*}(s) ] \nonumber
\end{eqnarray}
both $\sup_{0 \le s \le t}\sup_{0 \le u \le s}[A(u, s)- \alpha(s-u)]$ and $\sup_{0 \le s \le t}[A\otimes \beta(s) - A^{*}(s) ]$ are defined to depend on the arrival process $A$, which further makes them dependent on each other. Similar dependence can be found from (\ref{tc_e}), (\ref{tp_e}), (\ref{tb_e}) and (\ref{td_ea}). Such inherent dependence makes it difficult to directly obtain independent case results from (\ref{tc_e}), (\ref{to_e}), (\ref{tp_e}), (\ref{tb_e}) or (\ref{td_ea}).

In the following subsection, we introduce a new concept called {\em stochastic strict server} to help decouple the dependence found in (\ref{tc_e}), (\ref{to_e}), (\ref{tp_e}), (\ref{tb_e}) and (\ref{td_ea}). As a result, independent case analysis on properties (P.2) -- (P.5) can be further conducted.

\subsection{Stochastic Strict Server}

The definition of stochastic strict server is inspired by the following observation. Wireless channel is a typical stochastic server. A wireless channel typically operates in two states. If the channel is in ``good'' condition, data can be sent and received through it; if the channel is in ``bad'' condition due to some impairment, no data can be sent. 

Based on the observation, we use two processes to characterize the behavior of a stochastic server. These two processes are an {\em ideal service process} $\hat{S}(t)$ and an {\em impairment process} $I(t)$. Let us denote by $\hat{S}(s, t) \equiv \hat{S}(t) -\hat{S}(s)$ the amount of service that the server would have delivered in interval $[s, t)$ if there had been no service impairment in the interval, and $I(s, t)$ the amount of service, called {\em impaired service}, that cannot be delivered in the interval due to some impairment to the server. Particularly, the actually delivered service $S(t)$ to the input satisfies, for all $t \ge 0$, 
\begin{equation}
S(t) = \hat{S}(t) - I(t),
\end{equation}
with $\hat{S}(0)=0$, $I(0)=0$ by convention. It is clear that $\hat{S}$ and $I$ are in ${\mathcal{F}}$.

We now define {\em stochastic strict server} as follows: 

\begin{definition} \label{d_sss}
A server $S$ is said to be a {\em stochastic strict server} providing {\em stochastic strict service curve} $\hat{\beta}\in {\mathcal{F}}$ with {\em impairment process} ${I}$ to a flow iff during any backlogged period $[s,t)$, the output $A^{*}(s,t)$ of the flow from the server satisfies
\begin{equation}
A^{*}(s,t) \ge \hat{\beta}(t-s) - I(s, t).
\end{equation}
\end{definition}

Note that the definition of stochastic strict server implies that 
$$\hat{\beta}(0) \le 0.$$
Under the deterministic network calculus, a similar concept called {\em strict service curve} is defined and used \cite{NetCal}, which states that a strict server providing strict service curve $\hat{\beta}$ iff during any backlogged period $[s,t)$, $A^{*}(s,t) \ge \hat{\beta}(t-s)$. From Definition \ref{d_sss}, it is clear that if there is no impairment or $I(s,t)=0$ for all $0 \le s \le t$, a stochastic strict server becomes (deterministic) strict server providing strict service curve $\hat{\beta}$.

We show in Lemma \ref{l_ls} below that if the impairment process of a stochastic strict server has a m.b.c stochastic arrival curve, the server has a stochastic service curve. 

\begin{lemma}\label{l_ls}
Consider a stochastic strict server $S$ providing stochastic strict service curve $\hat{\beta}$ with impairment process ${I}$. If the impairment process has a m.b.c stochastic arrival curve, or $I \sim_{mb} \langle g, \gamma \rangle$, and $\beta \in {\mathcal{F}}$, then the server provides a stochastic service curve $S \sim_{sc} \langle g, \beta \rangle$ with 
$$\beta(t) = \hat{\beta}(t) - \gamma(t).$$
\end{lemma}

\begin{proof}
For any time $s \ge 0$, there are two cases. Case 1: $s$ is not within any backlogged period. In this case, there is no backlog in the server at $s$, which implies that all traffic arrived up to time $s$ has left the server. Hence, $A^{*}(s)=A(s)$ and consequently $A \otimes \beta(s)-A^{*}(s) \le A(s) + \beta(0) - A^{*}(s) \le 0$. Case 2: $s$ is within a backlogged period. Without loss of generality, assume the backlogged period starts from $s_0$. Then, $A^{*}(s_0)=A(s_0)$, and 
\begin{eqnarray}
&& A \otimes \beta(s)-A^{*}(s) \nonumber\\
&\le& A(s_0) + \beta(s-s_0) - A^{*}(s)  \nonumber \\
&=& \beta(s-s_0) + A^{*}(s_0)- A^{*}(s) = \beta(s-s_0) - A^{*}(s_0, s) \nonumber\\
&\le& I(s_0, s) + \beta(s-s_0) - \hat{\beta}(s-s_0)  \nonumber\\
&=& I(s_0, s) - \gamma(s-s_0) \nonumber\\
&\le& \sup_{0 \le u \le s}[I(u, s) - \gamma(s-u)]. \nonumber
\end{eqnarray}
Combining both cases, we conclude that for any $s \ge 0$,
\begin{eqnarray}
A \otimes \beta(s)-A^{*}(s) 
&\le& \left ( \sup_{0 \le u \le s}[I(u, s) - \gamma(s-u)] \right)^+ . \label{ls_1}
\end{eqnarray}
With (\ref{ls_1}), we further have 
\begin{eqnarray}
&& \sup_{0 \le s \le t}[A \otimes \beta(s)-A^{*}(s)] \nonumber \\
&\le& \left(\sup_{0\le s \le t}\sup_{o \le u \le s} [I(u, s) - \gamma(s-u)]\right)^+ \label{e_sss}
 %\nonumber \\&=& \left(\sup_{0\le u \le t}\sup_{u \le s \le t} [I(u, s) - \gamma(s-u)]\right)^+. \label{e_sss}
\end{eqnarray}
with which, $I \sim_{mb} \langle f, \gamma \rangle$, Lemma \ref{l_plus} and the definition of stochastic service curve, the proof is complete.
\end{proof}

\subsection{Properties (P.1) -- (P.5)} %Independent Case

With the concept of stochastic strict server and Lemma \ref{l_ls}, we now prove properties (P.1) -- (P.5) for the independent case. 

\begin{theorem} \label{ti_sup}
{\bf (P.1: Superposition)} Under the same condition as Theorem \ref{t_sup}, if $A_i(t)$, $i=1, \dots, N$, are independent, then $A \sim_{mb} \langle 1- \bar{f}_1 \ast \cdot \ast \bar{f}_i \cdot \ast \bar{f}_N, \sum_{i=1}^{N}\alpha_i \rangle$, where $\bar{f}_i(x) = 1-[f_i(x)]_1$, $i=1, 2, \dots, N$.
\end{theorem}
\begin{proof}
We only prove the case of $N=2$. For $N >2$, the proof can be conducted iteratively. 

We have proved (\ref{ts_e}) with which, since $A_1$ and $A_2$ are independent, the two terms on the right hand side of (\ref{ts_e}) are also independent. Then the theorem follows from Lemma \ref{l_plus} and Lemma \ref{l_rsumi}.
\end{proof}

\begin{theorem}\label{ti_con}
{\bf (P.2: Concatenation)} Under the same condition as Theorem \ref{t_con}, assume each node is a stochastic strict server providing stochastic strict service curve $\hat{\beta}^n$ with impairment process $I^n \sim_{mb} \langle g^n, \gamma^n \rangle$. If $I^n$ are independent and $\beta^n \in{\mathcal{F}}$, $(n = 1, 2, \dots, N)$, then the network guarantees to the flow a stochastic service curve $S \sim_{sc} \langle g, \beta \rangle$ with 
\begin{eqnarray}
\beta(t) &=& \beta^1 \otimes \beta^2 \otimes \cdots \otimes \beta^N (t) \nonumber \\
g (x) &=& 1- \bar{g}^1 \ast \bar{g}^2 \ast \cdots \ast \bar{g}^N (x), \nonumber 
\end{eqnarray}
where $\beta^n(t) = \hat{\beta}^n (t) - \gamma^n(t)$, $\bar{g}^n(x) = 1- [g^n(x)]_1$, $n = 1, 2, \dots, N$.
\end{theorem}
\begin{proof}
We only prove the two node case. For $N >2$, the proof can be extended easily.

In (\ref{tc_e}), we have proved
\begin{eqnarray}
&& \sup_{0 \le s \le t}[A\otimes \beta^1 \otimes \beta^2 (s) - A^*(s)] \nonumber \\
&\le& \sup_{0 \le s \le t}[A^1 \otimes \beta^1 (s) - A^{1*}(s)] \nonumber \\
&& + \sup_{0 \le s \le t}[A^2 \otimes \beta^2 (s)-A^{2*}(s)] \label{tc_ei}.
\end{eqnarray}
Then, with the stochastic strict server assumption for both servers, we can follow the same approach as in the proof of Lemma \ref{l_ls} to get (\ref{e_sss}) for them, applying which to (\ref{tc_ei}), we obtain
\begin{eqnarray}
&& \sup_{0 \le s \le t}[A\otimes \beta^1 \otimes \beta^2 (s) - A^*(s)] \nonumber \\
&\le& \left(\sup_{0\le s \le t}\sup_{0 \le u \le s} [I^1(u, s) - \gamma^1(s-u)]\right)^+ \nonumber \\
&& + \left(\sup_{0\le s \le t}\sup_{0 \le u \le s} [I^2(u, s) - \gamma^2(s-u)]\right)^+ \label{tc_eia}.
\end{eqnarray}
Since $I^1$ and $I^2$ are independent and so are the two terms of the right hand side of (\ref{tc_eia}), the theorem follows from Lemma \ref{l_plus} and Lemma \ref{l_rsumi}.
\end{proof}

\begin{theorem}\label{ti_out}
{\bf (P.3: Output Characterization)} Under the same condition as Theorem \ref{t_out}, assume the server is a stochastic strict server providing stochastic strict service curve $\hat{\beta}$ with impairment process $I\sim_{mb} \langle g, \gamma \rangle$. If $A$ and $I$ are independent, and $\beta \in {\mathcal{F}}$, then the output has a m.b.c stochastic arrival curve $A^{*} \sim_{mb} \langle f^{*}, \alpha^{*} \rangle$ with 
\begin{eqnarray}
\alpha^{*} (t) &=& \alpha \oslash \beta (t) \nonumber \\
f^{*}(x) &=& 1- \bar{f} \ast \bar{g} (x) \nonumber
\end{eqnarray}
where $\beta(t)=\hat{\beta}(t) - \gamma(t)$, $\bar{f}(x)=1-[f(x)]_1$ and $\bar{g}(x)=1-[g(x)]_1$.
\end{theorem}
\begin{proof}
In (\ref{to_e}), we have proved
\begin{eqnarray}
&& \sup_{0 \le s \le t}\sup_{0 \le u \le s}[A^{*}(u,s) - \alpha^{*}(s-u)] \nonumber \\
&\le& \sup_{0 \le s \le t}\sup_{0 \le u \le s}[A(u, s)- \alpha(s-u)] \nonumber \\
&& + \sup_{0 \le s \le t}[A\otimes \beta(s) - A^{*}(s) ]. \label{to_ei}
\end{eqnarray}
Then, with the stochastic strict server assumption, we can follow the same approach as in the proof of Lemma \ref{l_ls} to get (\ref{e_sss}). Applying (\ref{e_sss}) to (\ref{to_ei}), we further get
\begin{eqnarray}
&& \sup_{0 \le s \le t}\sup_{0 \le u \le s}[A^{*}(u,s) - \alpha^{*}(s-u)] \nonumber \\
&\le& \left( \sup_{0 \le s \le t}\sup_{0 \le u \le s}[A(u, s)- \alpha(s-u)]\right)^+ \nonumber \\
&& + \left(\sup_{0\le u \le t}\sup_{u \le s \le t} [I(u, s) - \gamma(s-u)]\right)^+ . \label{to_eia}
\end{eqnarray}
Sine $A$ and $I$ are independent and so are the first two terms on the right hand side of (\ref{to_eia}), the theorem follows easily from Lemma \ref{l_plus} and Lemma \ref{l_rsumi}.
\end{proof}

\begin{theorem}\label{ti_per}
{\bf (P.4: Per-Flow Service)} Similar to Theorem \ref{t_per}, consider a server fed with a flow $A$ that is the aggregation of two constituent flows $A_1$ and $A_2$. Assume the server is a stochastic strict server to the aggregate, providing stochastic strict service curve $\hat{\beta}$ with impairment process $I$.  
\begin{itemize}
\item[(i)] The server guarantees that
\begin{eqnarray}
&& \sup_{0 \le s \le t}[A_1 \otimes (\beta - \alpha_2)(s) - A^*_1(s)] \nonumber\\
&\le& \left(\sup_{0\le s \le t}\sup_{o \le u \le s} [I(u, s) - \alpha(s-u)\right)^+ \nonumber\\
&& + \left(\sup_{0 \le s \le t}\sup_{0 \le u \le s}[A_2(u,s) - \alpha_2(s-u)]\right)^+ \label{tp_eia}
\end{eqnarray}
\item[(ii)] If $A_2$ and $I$ are independent, $A_2 \sim_{mb} \langle f_2, \alpha_2 \rangle$,  $I\sim_{mb} \langle g, \gamma \rangle$, and $\beta'_1 \in {\mathcal{F}}$, then the server guarantees to flow $A_1$ a stochastic service curve $S_1 \sim_{sc} \langle g'_1, \beta'_1 \rangle$, where,
$$
g'_1(x) = 1- \bar{g}\ast \bar{f}_2(x); \quad \beta'_1(t)= \beta(t) - \alpha_2(t)
$$
with $\beta(t) = \hat{\beta}(t) - \gamma(t)$, $\bar{g}(x) = 1-[g(x)]_1$ and $\bar{f}_2(x) = 1- [f_2(x)]_1$.
\end{itemize}
\end{theorem}

\begin{proof}
For (i), we have proved in (\ref{tp_e}) that
\begin{eqnarray}
&& \sup_{0 \le s \le t}[A_1 \otimes (\beta - \alpha_2)(s) - A^*_1(s)] \nonumber\\
&\le& \sup_{0 \le s \le t}[A\otimes \beta(s) - A^*(s)] \nonumber\\
&& + \sup_{0 \le s \le t}\sup_{0 \le u \le s}[A_2(u,s) - \alpha_2(s-u)] \label{tp_ei}
\end{eqnarray}
Then with the stochastic strict server assumption, we can follow the same approach as in the proof of Lemma \ref{l_ls} to get (\ref{e_sss}) and apply it to (\ref{tp_ei}) to get (\ref{tp_eia}).

For (ii), since $I$ and $A_2$ are independent, the right hand side of (\ref{tp_eia}) are independent. Then the second part (ii) follows from Lemma \ref{l_plus} and Lemma \ref{l_rsumi}.
\end{proof}

%\begin{remark}
{\bf Remark:} (i) From the proof, it is clear that (\ref{tp_eia}) is an intermediate step for getting Theorem \ref{ti_per}.(ii). The intention of including (\ref{tp_eia}) as part of the theorem is as follows:  When  (P.4) is used to derive other results such as (P.2), (P.3) and (P.5), Theorem \ref{ti_per}.(i) or (\ref{tp_eia}) can be applied to their derivations, e.g. (\ref{tc_ei}) for (P.2), (\ref{to_ei}) for (P.3), and (\ref{tb_ei}) and (\ref{td_ei}) for (P.5). Then, if flows and the impairment processes of servers are independent, Lemma \ref{l_rsumi} can be used to derive the corresponding independent case bounds. However, if we were only given Theorem \ref{ti_per}.(ii), such independent case analysis could not be applied and the general case (min, +) analysis in the previous section would have to be used. As a result, looser bounds may be obtained. 

(ii) From the view point of the service provided to flow $A_1$, $A_2(t)$ can be considered as an impairment process. In other words, to flow $A_1$, the server has two independent impairment processes $I(t)$ and $A_2(t)$. With this view point, the theorem can also be proved based on Theorem \ref{ti_sup} and Lemma \ref{l_ls}.
%\end{remark}

\begin{theorem}\label{ti_ser}
{\bf (P.5: Service Guarantees)} Under the same condition as Theorem \ref{t_ser}, if the server is a stochastic strict server providing stochastic strict service curve $\hat{\beta}$ with impairment process $I\sim_{mb} \langle g, \alpha \rangle$, and if $A$ and $I$ are independent, then 
\begin{itemize}
\item[(i)] The backlog $B(t)$ is guaranteed that for all $x\ge0$, $
P \{B(t) > x\} \le 1 - \bar{f}\ast \bar{g}(x+ \inf_{s \ge 0}[\beta(s) - \alpha(s)])
$
\item[(ii)] The delay $D(t)$ is guaranteed that for all $x \ge 0$, $
P \{D(t) > x\} \le 1- \bar{f} \ast \bar{g}(\inf_{s \ge 0}[\beta(s) - \alpha(s-x)]),
$
\end{itemize}
where $\bar{f}(x)=1-[f(x)]_1$ and $\bar{g}(x) = 1-[g(x)]_1$.
\end{theorem}
\begin{proof}
For the backlog $B(t)$, we have proved in (\ref{tb_e}) that 
\begin{eqnarray}
&& B(t) \nonumber\\
&\le& \sup_{0 \le s \le t}\sup_{0 \le u \le s}[A(u,s) - \alpha(s-u)] \nonumber\\
&& + \sup_{0 \le s \le t}[A\otimes \beta(t) - A^*(t)] - \inf_{s \ge 0}[\beta(s) - \alpha(s)] \label{tb_ei}.
\end{eqnarray}
Then, with the stochastic strict server assumption for both servers, we can follow the same approach as in the proof of Lemma \ref{l_ls} to get (\ref{e_sss}) for them, applying which to (\ref{tb_ei}), we obtain
\begin{eqnarray}
&& B(t) \nonumber\\
&\le& \left(\sup_{0 \le s \le t}\sup_{0 \le u \le s}[A(u,s) - \alpha(s-u)]\right)^+ \nonumber\\
&&+ \left(\sup_{0\le s \le t}\sup_{0 \le u \le s} [I(u, s) - \gamma(s-u)]\right)^+ \nonumber \\
&& - \inf_{s \ge 0}[\beta(s) - \alpha(s)] \label{tb_eia}.
\end{eqnarray}
Since $A$ and $I$ are independent and so are the first two terms on the right hand side of (\ref{tb_eia}), the first part follows from Lemma \ref{l_plus} and Lemma \ref{l_rsumi}.

For the delay $D(t)$, we have proved in (\ref{td_e})
\begin{eqnarray}
&& P\{D(t) > x\} \nonumber\\
&\le& P \{ X_a + Y_s > \inf_{s \ge 0}[\beta(s) - \alpha(s-x)]\} \label{td_ei}
\end{eqnarray}
where $X_a = \sup_{0 \le s \le t}\sup_{0 \le u \le s}[A(u,s) - \alpha(s-u)]$ and $Y_s = \sup_{0 \le s \le t+x}[A\otimes \beta(s) - A^*(s)]$. Based on the stochastic strict server assumption and independence assumption, we can follow similar approach above for backlog and obtain
\begin{eqnarray}
&& P\{D(t) > x\} \nonumber\\
&\le& P \{ X'_a + Y'_s > \inf_{s \ge 0}[\beta(s+x) - \alpha(s)]\} \label{td_eia}
\end{eqnarray}
where 
\begin{eqnarray}
X'_a &=& \left(\sup_{0 \le s \le t}\sup_{0 \le u \le s}[A(u,s) - \alpha(s-u)] \right)^+  \nonumber\\
Y'_s &=& \left(\sup_{0\le s \le t+x}\sup_{0 \le u \le s} [I(u, s) - \gamma(s-u)]\right)^+. \nonumber
\end{eqnarray}
Since $A$ and $I$ are independent, $X'_a$ and $Y'_s$ are independent. The delay part then follows from Lemma \ref{l_plus} and Lemma \ref{l_rsumi}.
\end{proof}

\section{Processes with m.b.c Stochastic Arrival Curve} \label{sec_5}

In the previous two sections, we have studied properties (P.1) -- (P.5) based on the concepts of m.b.c stochastic arrival curve, stochastic service curve and stochastic strict server. Note that stochastic strict server can be used to find the stochastic service curve of a server and the impairment process of stochastic strict server can also be characterized using m.b.c stochastic arrival curve. Viewing these, we focus this section on deriving results that can help find the m.b.c stochastic arrival curve of a process.

For ease of exposition, let us denote by $M(t)$ the maximum up-to-date backlog at time $t$ in a initially empty system with constant service rate $r$ and arrival process $A(t)$, or formally
\begin{equation} \label{e_mb}
M(t) \equiv \sup_{0\le s \le t}\sup_{0 \le u \le s} [A(u,s) - r(s-u)].
\end{equation}
It can be easily verified that, for $M(t)$, the following equation holds
\begin{equation} \label{e_mb1}
M(t+1) = max[M(t), W(t+1)]
\end{equation}
where
$$
W(t+1) = \sup_{0 \le s \le t+1} [A(s, t+1) - r(t+1-s)].
$$
With (\ref{e_mb1}), it is clear that
\begin{eqnarray}
M(t) \le M(t+1) \le \cdots \le M(\infty). \label{e_mbod}
\end{eqnarray}

Equations (\ref{e_mb}) and (\ref{e_mbod}) present the basic technique for finding the m.b.c stochastic arrival curve characterization of a process. They imply that if for some $r$ the compliment distribution $P\{M(t) >x\}$ can be found, then the process has $A \sim_{mb} \langle P\{M(t) >x\}, rt \rangle$. In addition, if $P\{M(\infty) >x\}$ exists, then it can also be used as the bounding function and we can conclude $A \sim_{mb} \langle P\{M(\infty) >x\}, rt \rangle$. 

However, (\ref{e_mbod}) shows that $M(t)$ increases over $t$, which implies that in general, $M(\infty)$ may not be bounded above by a constant, and $M(t)$ could not have a limit distribution. Because of this, additional conditions on the process may be needed to ensure that $M(t)$ converges in distribution. Such conditions have been discussed in the literature for the study of  maximum queue length \cite{Anderson70}\cite{Serfozo88}\cite{SS92}. In the rest, we consider a condition that has been widely used (e.g \cite{Chang94}\cite{Chang00}).

The condition is that $a(t)$ is a sequence of independent identically distributed (i.i.d.) random variables. Here, $a(t)$ denotes $a(t) \equiv A(t) - A(t-1)$ for all $t \ge 1$ with $a(0)=0$. Intuitively, for the arrival process, $a(t)$ represents the amount of traffic arriving at time $t$. Under this condition, the following result is important.

\begin{lemma} \label{l_mbst}
If $a(t)$ is i.i.d., there holds
\begin{eqnarray}
M(t) &\le_{st}& \sup_{t \ge 0} [A(t) - r \cdot t]. \label{e_mbst2}
\end{eqnarray}
\end{lemma}

\begin{proof}
Since $a(t)$ is i.i.d., $A(s, t)$ is stationary and the following relationship holds:
\begin{eqnarray}
M(t) &\equiv& \sup_{0\le s \le t}\sup_{0 \le u \le s} [A(u, s) - r(s-u)] \nonumber\\
&=_{st}& \sup_{0\le s \le t}\sup_{0 \le u \le s} [A(s-u) - r(s-u)]. \label{e_mbst}
\end{eqnarray}
While (\ref{e_mbst}) seems to be obvious since $A(u,s)$ is stationary, it actually follows from a non-trivial result on stochastic ordering, which is Theorem 2.2.3 of \cite{Stoyan83}.

The right hand side of (\ref{e_mbst}) can be simplified as:
\begin{eqnarray}
&& \sup_{0\le s \le t}\sup_{0 \le u \le s} [A(s-u) - r(s-u)] \nonumber\\
&=& \sup_{0\le s \le t}\sup_{0 \le v \le s} [A(v) - r \cdot v] \nonumber\\
&\le& \sup_{0\le s \le t}\sup_{v \ge 0} [A(v) - r \cdot v] \nonumber\\
&=& \sup_{v \ge 0} [A(v) - r \cdot v] \label{e_mbst1}
\end{eqnarray}
with which and (\ref{e_mbst}), (\ref{e_mbst2}) is proved.
\end{proof}

Note that when $A(t)$ is the superposition or aggregation of multiple independent progresses $A_i(t), (i=1, 2, \dots, n)$ and each of them is i.i.d. associated with a rate parameter $r_i$, step (\ref{e_mbst}) can be rewritten as 
\begin{eqnarray}
M(t) &=& \sup_{0\le s \le t}\sup_{0 \le u \le s} [\sum_{i=1}^{n}A_i(u, s) - \sum_{i=1}^{n}r_i(s-u)] \nonumber\\
&=_{st}& \sup_{0\le s \le t}\sup_{0 \le u \le s} [\sum_{i=1}^{n}A_i(s-u) - \sum_{i=1}^{n}r_i(s-u)]\nonumber
\end{eqnarray}
from which, (\ref{e_mbst2}) also follows, or, 
\begin{eqnarray}
M(t) &\le_{st}& \sup_{t \ge 0} [A(t) - r \cdot t],
\end{eqnarray}
where $A(t)=\sum_{i=1}^{n}A_i(s-u)$ and $r=\sum_{i=1}^{n}r_i$.

With Lemma \ref{l_mbst}, we now relate the $(\sigma(\theta), \rho(\theta))$ characterization \cite{Chang94} \cite{Chang96} \cite{Chang00} to the m.b.c stochastic arrival curve characterization of a process. 

A process $A(t)$ is said to be $(\sigma(\theta), \rho(\theta))$-upper constrained (for some $\theta > 0$), iff for all $0 \le s \le t$ \cite{Chang94}\cite{Chang00}
\begin{equation} \label{e_srth}
\frac{1}{\theta}log E e^{\theta A(s,t)} \le \rho(\theta) (t-s) + \sigma(\theta).
\end{equation}
It has been shown that if $a(t)$ is i.i.d. or has a stationary Markov modulated process (MMP) or some other process, $A(t)(=\sum_{s=0}^{t}a(s))$ can be represented using the $(\sigma(\theta), \rho(\theta))$ characterization \cite{Chang94}\cite{Chang96}\cite{Chang00}.

\begin{theorem}\label{t_trc}
If $a(t), t \ge 0$, is i.i.d. and the corresponding $A(t)$ is $(\sigma(\theta), \rho(\theta))$-upper constrained, then it has a m.b.c stochastic arrival curve $A \sim_{mb} \langle f, \alpha \rangle$, where
\begin{eqnarray}
\alpha(t) &=& r \cdot t \nonumber \\
f(x) &=& \frac{e^{\theta \sigma(\theta)}}{1-e^{\theta(\rho(\theta)-r)}}e^{-\theta x} \nonumber
\end{eqnarray}
for any $r > \rho(\theta)$.
\end{theorem}

\begin{proof}
For ease of exposition, let $W \equiv \sup_{v \ge 0} [A(v) - r \cdot v]$. Since $A(t)$ is $(\sigma(\theta), \rho(\theta))$-upper constrained, we shall prove that for all $x \ge 0$,
\begin{equation}
P\{W > x\} \le f(x), \label{e_mbst3}
\end{equation}
from which and Lemma \ref{l_mbst}, the theorem follows easily.

To prove (\ref{e_mbst3}), let us consider $e^{\theta W}$. Since for any $\theta \ge 0$, $e^{\theta x}$ is increasing in $x$, we then obtain
\begin{eqnarray}
e^{\theta W} &\le& \sup_{v \ge 0} \left [ e^{ \theta ( A(v) - r \cdot v)} \right ] \nonumber\\
&\le& \sum_{v=0}^{\infty}\left [ e^{ \theta ( A(v) - r \cdot v)} \right ],
\end{eqnarray}
where the second step follows from the inequality that $\sup \{x, y\} \le x+y$ for any $x, y \ge 0$.

Since $A$ is $(\sigma(\theta),\rho(\theta))$-upper constrained, or, $E e^{ \theta A(v)} \le e^{\theta \rho(\theta) v + \theta \sigma(\theta)}$ for all $v \ge 0$, then we have, 
\begin{eqnarray}
E e^{\theta W} &\le& \sum_{u=0}^{\infty}\left [ E e^{ \theta ( A(v) - r \cdot v)} \right ] \nonumber \\
&\le& \sum_{u=0}^{\infty}\left [ e^{ \theta (\rho(\theta) - r)v + \theta \sigma(\theta) } \right ] \nonumber \\
& = & \frac{e^{\theta \sigma(\theta)}}{1-e^{\theta(\rho(\theta)-r)}}. \label{ew}
\end{eqnarray}

Finally, (\ref{e_mbst3}) is obtained from Chernoff's bound for random variable, or 
$P\{W > x\} = P\{e^{\theta W} > e^{\theta x}\} \le f(x)$.
\end{proof}

Theorem \ref{t_trc} implies that if the $(\sigma(\theta), \rho(\theta))$-characterization of a process is known, its m.b.c stochastic arrival curve could be derived. In addition, if the $(\sigma(\theta), \rho(\theta))$-characterization of the impairment process in a stochastic strict server is known, the stochastic service curve of the server could also be obtained from Theorem \ref{t_trc}.

\section{Conclusions} \label{sec_6}

This paper has made several contributions in the context of stochastic network calculus. First, we defined two new generalizations of arrival curve and service curve, which are respectively {\em m.b.c stochastic arrival curve} and {\em stochastic service curve}. With them, we have been able to apply the (min, +) analysis to prove the five basic properties (P.1) -- (P.5) for the stochastic network calculus without any additional constraint on the bounding functions. Second, we introduced a new notion of server characterization: {\em stochastic strict server}. The idea of stochastic strict server is to decouple the service process into an {\em ideal service process} and a service {\em impairment process}. Importantly, with such decoupling, we have conducted independent case analysis and proved for the first time properties (P.1) -- (P.5) for the independent case. Third, we proved that if a process is i.i.d. and is $(\sigma(\theta), \rho(\theta))$-upper constrained, it has a m.b.c stochastic arrival curve. Note that with stochastic strict server, the stochastic service curve of a server is found when the m.b.c stochastic arrival curve of the impairment process is known. Since $(\sigma(\theta), \rho(\theta))$ characterization can be used to represent many well-known processes, we hence believe our results are useful no only in finding the m.b.c stochastic arrival curve of a flow but also in helping find the stochastic service curve of a server. 

It can be easily verified that the corresponding properties (P.1) -- (P.5) of the (min, +) deterministic network calculus can be recovered from the basic stochastic network calculus developed in this paper. A future work is to apply the basic calculus to study other problems (e.g. routing \cite{Chang00} and feedback control \cite{Agrawal99}) that have been addressed by the deterministic network calculus but yet for the stochastic network case. Another future work is to extend the calculus to analyze quality of service guarantees in wireless networks.

\section*{Acknowledgment} 
We thank Cheng-Shang Chang for pointing out a mistake in Theorem \ref{t_trc} and its proof in an early version of the paper.

%\bibliographystyle {abbrv} %{unsrt} %{abbrv}
%\bibliography{netcal}  % the name of the Bibliography

\end{document}